\documentclass[english,twocolumn,showpacs,preprintnumbers,amsmath,amssymb,aps,dblfloatfix]{revtex4-1}
\usepackage[T1]{fontenc}
\usepackage[latin9]{inputenc}
\usepackage{color}
\usepackage{array}
\usepackage{units}
\usepackage{multirow}
\usepackage{amsmath}
\usepackage{amssymb}
\usepackage{graphicx}

\makeatletter

\providecommand{\tabularnewline}{\\}

\@ifundefined{textcolor}{}
{%
 \definecolor{BLACK}{gray}{0}
 \definecolor{WHITE}{gray}{1}
 \definecolor{RED}{rgb}{1,0,0}
 \definecolor{GREEN}{rgb}{0,1,0}
 \definecolor{BLUE}{rgb}{0,0,1}
 \definecolor{CYAN}{cmyk}{1,0,0,0}
 \definecolor{MAGENTA}{cmyk}{0,1,0,0}
 \definecolor{YELLOW}{cmyk}{0,0,1,0}
}

\usepackage{lmodern}
\usepackage[T1]{fontenc}
\usepackage{mathtools}
\usepackage{url}
\usepackage{ragged2e}
\edef\UrlBreaks{\do\-\UrlBreaks}

\makeatother

\usepackage{babel}
\begin{document}

\title{Analytic potentials and vibrational energies for Li$_{2}$ states
dissociating to $\mbox{Li}\left(2S\right)+\mbox{Li}\left(3P\right)$.
Part 1: The $^{2S+1}\Pi_{\nicefrac{u}{g}}$ states}

\author{Nikesh S. Dattani,}

\email{dattani.nike@gmail.com }

\affiliation{}

\affiliation{$^{1}$School of Materials Science and Engineering, Nanyang Technological
University, 639798, Singapore,}

\affiliation{$^{2}$Fukui Institute for Fundamental Chemistry, 606-8103, Kyoto,
Japan,}

\affiliation{$^{3}$Quantum Chemistry Laboratory, Department of Chemistry, Kyoto
University, 606-8502, Kyoto, Japan,}

\author{}
\begin{abstract}
Analytic potentials are built for all four $^{2S+1}\Pi_{u/g}$ states
of Li$_{2}$ dissociating to Li$(2S)$ + Li$(3P)$: $3b(3^{3}\Pi_{u})$,
$3B(3^{1}\Pi_{u})$, $3C(3^{1}\Pi_{g}),$ and $3d(3^{3}\Pi_{g})$.
These potentials include the effect of spin-orbit coupling for large
internuclear distances, and include state of the art long-range constants.
This is the first successful demonstration of fully analytic diatomic
potentials that capture features that are usually considered too difficult
to capture without a point-wise potential, such as multiple minima,
and shelves. Vibrational energies for each potential are presented
for the isotopologues $^{6,6}$Li$_{2}$, $^{6,7}$Li$_{2}$, $^{7,7}$Li$_{2}$,
and the elusive `halo nucleonic molecule' $^{11,11}$Li$_{2}$. These
energies are claimed to be accurate enough for new high-precision
experimental setups such as the one presented in {[}Sebastian \emph{et
al.} Phys. Rev. A, \textbf{90}, 033417 (2014){]} to measure and assign
energy levels of these electronic states, all of which have not yet
been explored in the long-range region. Measuring energies in the
long-range region of these electronic states may be significant for
studying the \emph{ab initio} vs experiment discrepancy discussed
in {[}Tang \emph{et al.} Phys. Rev. A, \textbf{84}, 052502 (2014){]}
for the $C_{3}$ long-range constant of Lithium, which has significance
for improving the SI definition of the second.
\end{abstract}


\maketitle
Very little is known about the Li$_{2}$ electronic states dissociating
to the $2S+3P$ asymptote. Out of the first 5 asymptotes (from lowest
to highest: $2S+2S$, $2S+2P$, $2S+3S$, $2P+2P$ and $2S+3P$),
the $2S+3P$ is the only one for which an empirical dissociation energy
has not been determined for any of the electronic states dissociating
to it. Furthermore, the only measurements that have been made for
electronic states dissociating to $2S+3P$, were for the $3c(3^{3}\Sigma_{g}^{+})$
state \cite{Xie1986,Yiannopoulou1995,Li2007} the $6X(6^{1}\Sigma_{g}^{+})$
state \cite{Bernheim1983,Song2002}, and the $3d(3^{3}\Pi_{g})$ state
\cite{Yiannopoulou1995a,Li1996a,Li1996,Ivanov1999}. No measurements
have been done on the other states dissociating to $2S+3P$.

Very recently, a promising experiment has been setup with the ability
to use photoassociation in a magneto-optical trap to make ultra-cold
$^{6}$Li$_{2}$ molecules dissociating to the $2S+3P$ asymptote
\cite{Sebastian2014}, much like slightly earlier experiments which
have already been successful for creating ultra-cold $^{6}$Li$_{2}$
molecules dissociating to $2S+2P$ with very similar techniques \cite{Semczuk2013,Gunton2013}.
Measurements of the binding energies for levels very close to the
$2S+3P$ asymptote would allow for an empirical determination of the
long-range constant $C_{3}^{2S+3P}$ which is the leading interaction
constant in the potential energy between Li$(2S)$ and Li$(3P)$.

\begin{table*}
\begin{tabular}{cc>{\raggedright}p{0.2cm}ccc>{\raggedright}p{2cm}}
\multirow{4}{*}{$u^{3b}(r)$} & \multirow{4}{*}{=} & \multirow{4}{0.2cm}{$\begin{cases}
\\
\\
\\
\\
\end{cases}$} & $u^{3b,0_{u}^{+}}(r)$ & $\xrightarrow{\qquad u^{6A,0_{u}^{+}}(r)\qquad}$ & $2S_{1/2}+3P_{1/2}$ & (higher) \label{3b0u+}\tabularnewline[-2mm]
 &  &  & $u^{3b,0_{u}^{-}}(r)$ & $\xrightarrow{\qquad u^{6a,0_{u}^{-}}(r)\qquad}$ & $2S_{1/2}+3P_{3/2}$ & (lower)\label{3b0u-}\tabularnewline
 &  &  & $u^{3b,1_{u}}(r)$ & $\xrightarrow{\,\, u^{3B,1_{u}}(r),u^{6a,1_{u}}(r)\,\,}$ & $2S_{1/2}+3P_{3/2}$ & (lowest)\label{3b1u}\tabularnewline
 &  &  & $u^{3b,2_{u}}(r)$ & $\xrightarrow{\qquad\qquad\qquad\qquad}$ & $2S_{1/2}+3P_{3/2}$ & \label{3b2u}\tabularnewline[6mm]
\hline 
\noalign{\vskip6mm}
$u^{3B}(r)$ & = &  & $u^{3B,1_{u}}(r)$ & $\xrightarrow{\,\, u^{6a,1_{u}}(r),u^{3b,1_{u}}(r)\,\,}$ & $2S_{1/2}+3P_{3/2}$ & (middle)\label{3B1u}\tabularnewline[6mm]
\hline 
\noalign{\vskip6mm}
$u^{3C}(r)$ & = &  & $u^{3C,1_{g}}(r)$ & $\xrightarrow{\,\, u^{3c,1_{g}}(r),u^{3d,1_{g}}(r)\,\,}$ & $2S_{1/2}+3P_{3/2}$ & (middle)\label{3C1g}\tabularnewline[6mm]
\hline 
\noalign{\vskip6mm}
\multirow{4}{*}{$u^{3d}(r)$} & \multirow{4}{*}{=} & \multirow{4}{0.2cm}{$\begin{cases}
\\
\\
\\
\\
\end{cases}$} & $u^{3d,0_{g}^{+}}(r)$ & $\xrightarrow{\qquad u^{6X,0_{g}^{+}}(r)\qquad}$ & $2S_{1/2}+3P_{1/2}$ & (lower)\label{3d0g+}\tabularnewline[-2mm]
 &  &  & $u^{3d,0_{g}^{-}}(r)$ & $\xrightarrow{\qquad u^{3c,0_{g}^{-}}(r)\qquad}$ & $2S_{1/2}+3P_{3/2}$ & (higher)\label{3d0g-}\tabularnewline
 &  &  & $u^{3d,1_{g}}(r)$ & $\xrightarrow{\,\, u^{3c,1_{g}}(r),u^{3C,1_{g}}(r)\,\,}$ & $2S_{1/2}+3P_{3/2}$ & (highest)\label{3d1g}\tabularnewline
 &  &  & $u^{3d,2_{g}}(r)$ & $\xrightarrow{\qquad\qquad\qquad\qquad}$ & $2S_{1/2}+3P_{3/2}$ & \label{3d2g}\tabularnewline[6mm]
\end{tabular}
\end{table*}

At the lower asymptote of $2S+2P$, there is a discrepancy between
experiment and theory for the long-range constant $C_{3}^{2S+3P}$,
despite Li only having 3e$^{-}$ and the experimental value being
the most precisely determined oscillator strength ever determined
for a molecule, by an order of magnitude \cite{Tang2011}. This has
various consequences, reaching as far as limiting progress towards
improving the precision of the SI definition of the second \cite{Mitroy2010}.
More precise atomic clocks are needed for various applications. The
current definition of the second is based on a clock transition frequency
in Cs with a relative uncertainty of $\sim5\times10^{-16}$, and a
commonly quoted target for improved precision is $10^{-18}$ \cite{Mitroy2010}.
The largest source of uncertainty limiting atomic clock precision
is the blackbody radiation shift, which depends on the static dipole
polarizability of the system being used for the atomic clock \cite{Mitroy2010}.
Lithium is expected to play a major role in polarizability metrology,
since polarizability ratios can be measured much more precisely than
individual polarizabilities \cite{Cronin2009} and Li is the preferred
choice for the standard in the denominator of such a ratio \cite{Mitroy2010}.
The discrepancy in $C_{3}$ limits the accuracy of a potential Li-based
standard for polarizabilities \cite{Tang2011}, and hence indirectly
impacts progress towards improving the SI definition of the second.

Regarding the empirical value for $C_{3}^{2S+2P}$, for most electronic
states, the mixing of various states towards the $2S+2P$ asymptote
significantly complicates the expressions from which $C_{3}$ is fitted
\cite{Dattani2015a,Dattani2011,LeRoy2009,Aubert-Frecon1998,Bussery1985}.
The complicated expressions for this mixing are the same at the $2S+3P$
asymptote as they are for the $2S+2P$ asymptote \cite{Aubert-Frecon2015},
but the fine structure splitting parameter which governs the significance
of this mixing, is about 3.5 times smaller at the $2S+3P$ asymptote
than at the $2S+2P$ asymptote. For $2S+2P$ the fine structure splitting
parameter for $^{6}$Li is $\Delta E_{2^{2}P_{\nicefrac{3}{2}}\leftarrow2^{2}P_{\nicefrac{1}{2}}}=D_{2}-D_{1}=0.335\,324\,6$~cm$^{-1}$
\cite{Semczuk2013,Gunton2013,Brown2013a,Sansonetti2011}, while for
$2S+3P$ it is only $\Delta E_{3^{2}P_{\nicefrac{3}{2}}\leftarrow3^{2}P_{\nicefrac{1}{2}}}=0.096$
cm$^{-1}$ \cite{Sansonetti1995}. Therefore, $C_{3}^{2S+3P}$ might
be a better benchmark for an \emph{ab initio }vs experiment comparison
than $C_{3}^{2S+2P}$, as the effect of this complication is smaller. 

Measuring and assigning molecular energy levels using photoassociation
requires reasonably accurate predictions which come from eigenvalues
of a Schr\"{o}dinger equation, and hence require a reasonably accurate
potential energy surface. Due to the shortage of measurements on the
Li$_{2}$ $^{2S+1}\Pi_{\nicefrac{u}{g}}$ states dissociating to $2S+3P$,
the most accurate potentials come from purely \emph{ab initio} calculations.
For the $3c(3^{3}\Sigma_{g}^{+})$ and $3A(3^{1}\Sigma_{u}^{+})$
states, \emph{ab initio} calculations were reported in 1985 \cite{Schmidt-Mink1985},
1995 \cite{Poteau1995}, 2006 \cite{Jasik2006} and 2014 \cite{Musia2014,Lee2014};
and for the $3d(3^{3}\Pi_{g})$ state in 1995 \cite{Yiannopoulou1995a}
and 2014 \cite{Musia2014,Lee2014}. But for the rest of the states
dissociating to $2S+3P$; namely $3b(3^{3}\Pi_{u})$, $6X(6^{1}\Sigma_{g}^{+})$,
$3B(3^{1}\Pi_{u})$, $3C(3^{1}\Pi_{g})$, and $6a(6^{3}\Sigma_{u}^{+})$;
the only \emph{ab initio} calculations reported were in \cite{Poteau1995,Musia2014,Lee2014}.
All of these \emph{ab initio} papers also reported potentials for
states dissociating to lower asymptotes, where plenty of experimental
data is available to gauge the quality of the calculations\@.

In my very recent paper on comparing experiment to \emph{ab initio}
for the $b(1^{3}\Pi_{u})$ state \cite{Dattani2015a}, it was found
that the \emph{ab initio} potential of \cite{Musia2014} predicted
all vibrational binding energies with a disagreement of $<12$~cm$^{-1}$
with the corresponding energies of the empirical potential. Furthermore
there was always $<0.8$~cm$^{-1}$ disagreement between the empirical
and \emph{ab initio} vibrational energy spacings. Finally, when comparing
the dissociation energies $\mathfrak{\mathfrak{D}}_{e}$ from \cite{Musia2014}
to the corresponding experimental values for all states which have
empirical $\mathfrak{D}_{e}$ values available, the \emph{ab initio
}values from \cite{Musia2014} were never in disagreement by more
than 68~cm$^{-1}$. Therefore, the \emph{ab initio} potentials from
\cite{Musia2014} for the states dissociating to $2S+3P$ are expected
to be a good starting point for predicting energy levels with the
precision required for photoassociation experiments as in \cite{Semczuk2013,Gunton2013}
and as may be preformed with the new setup in \cite{Sebastian2014}
which is capable of detecting states dissociating to $2S+3P$.

However, the \emph{ab initio }calculations of \cite{Musia2014} still
have some major drawbacks (including, but not limited to):
\begin{enumerate}
\item the \emph{ab initio} points are not on a dense enough mesh to use
as the mesh for solving the effective radial Schr\"{o}dinger equation
for predicting the vibrational energies (\emph{especially} for large
distances where the energies become more important for fitting an
empirical $C_{3}$ value, and for experiments such as those potentially
resulting from a setup such as in \cite{Sebastian2014});
\item the \emph{ab initio} points neglect the effect of spin-orbit coupling,
which is particularly important for $nS+n^{\prime}P$ states of Li$_{2}$,
where the effect of interstate coupling has been shown to be absolutely
obligatory for describing the high vibrational energy measurements
\cite{LeRoy2009,Dattani2011,Semczuk2013,Gunton2013};
\item the \emph{ab initio} points do not go beyond the Born-Oppenheimer
approximation (so do not distinguish between different isotopologues
such as $^{6,6}$Li$_{2},$ $^{6,7}$Li$_{2}$, $^{7,7}$Li$_{2}$,
and the elusive `halo nucleonic' isotopologues containing $^{11}$Li),
and they are non-relativistic. 
\end{enumerate}
Drawback (1) is usually treated by fitting an interpolant through
the \emph{ab initio} points, but the resulting energies predicted
after solving the Schr\"{o}dinger equation, will be very sensitive to
the type of interpolant used, especially at the level of precision
of photoassociation experiments (the precision in the Li$_{2}$ measurements
of \cite{Semczuk2013,Gunton2013} was $\pm0.00002$~cm$^{-1}$ or
$\pm600$~kHz). Also, if for sake of ease, a spline interpolant is
used, it would be defined piecewise and would have discontinuous first
derivatives. The spline also knows nothing about the physics of nature,
and will therefore not know what to do in regions where fewer \emph{ab
initio} points are available (in this example, for large internuclear
distances). 

Rather than interpolating with a spline, we can fit to a fully analytic
model potential that has the correct theoretical behavior incorporated
in the long-range region where fewer \emph{ab initio} points are available,
and this addresses drawback (2) as well, since the effect of spin-orbit
coupling at long-range can easily be incorporated into the model.
Part of drawback (3) can also be addressed by fitting to a model potential,
because the model can also build in some types of relativistic effects
such as QED retardation, as was attempted in \cite{LeRoy2009,Dattani2011,Semczuk2013,Gunton2013}.
In 2011 the Morse/long-range (MLR) potential was fitted to spectroscopic
data for the $c(1^{3}\Sigma_{g}^{+})$ state of Li$_{2}$ where there
was a gap of $>5000$~cm$^{-1}$ between data near the bottom of
the potential's well, and data at the very top \cite{Dattani2011}.
In 2013 it was found by experiment that vibrational energies predicted
from this MLR potential in the very middle of this gap were correct
to about 1~cm$^{-1}$ \cite{Semczuk2013}. Therefore, fitting the
\emph{ab initio} data to the MLR model can provide reliable energy
predictions in regions where \emph{ab initio} points are lacking or
are poor in quality.

Therefore, in this paper MLR models that incorporate the long-range
theoretical effect of spin-orbit coupling are fitted to the \emph{ab
initio} points from \cite{Musia2014} for the $^{2S+1}\Pi_{\nicefrac{u}{g}}$
states of Li$_{2}$ dissociating to $2S+3P$. Drawback (3) is not
addressed in this paper. However, Born-Oppenheimer breakdown (BOB)
corrections could have been added to the \emph{ab initio }points using
the molecular electron wavefunction as described in \cite{Dattani2015a}.
Alternatively, the entire \emph{ab initio} calculation can be redone
using a non-Born-Oppenheimer approach as has been done for up to 6e$^{-}$
\cite{Bubin2009a}, but the \emph{a posteriori} approach of doing
a Born-Oppenheimer calculation and then adding BOB corrections afterwards
has been shown to work better according to the agreement between experiment
and theory for BeH \cite{Dattani2015,Bubin2007}. Also, DKH (Douglas-Kroll-Hess)
relativistic corrections can be added to the \emph{ab initio} potential
as in \cite{Koput2011}, and QED effects can also be added as was
done for H$_{2}$ in \cite{Komasa2011} and HeH$^{+}$ in \cite{Pachucki2012}.
If any of these answers to drawback (3) were to be addressed by adding
corrections to the \emph{ab initio} points of \cite{Musia2014}, the
procedure applied in the present paper for fitting an MLR function
to \emph{ab initio} points, could be repeated for even more accurate
analytic potentials.

\section{Extending the \emph{ab initio} calculations}

The \emph{ab initio }calculations in \cite{Musia2014} did not go
beyond 22~\textcolor{black}{$\mbox{\AA}$. However, beyond a certain
length, analytic expressions for the potential can be derived from
the theory of atom-atom interactions with disregard for the effect
of overlap between each atom's electronic wavefunction. These analytic
expressions are based on long-range constants that come from atomic
}\textcolor{black}{\emph{ab initio}}\textcolor{black}{{} calculations
rather than molecular ones, so for Li$_{2}$ the calculations only
involve 3e$^{-}$ rather than 6e$^{-}$. This means, for example,
that a coupled cluster calculation taking account of the full configuration
interaction (FCI) of a basis set only needs to go up to triple excitations
(CCSDT, whose scaling with respect to the number of basis functions
$N$ is $\sim N^{8}$ and has been implemented since 1987 \cite{Noga1987});
whereas a molecular calculation on Li$_{2}$ would require all the
way up to hexuple excitations (CCSDTQPH, which scales as $\sim N^{14}$,
and has been implemented in only very few studies since 2000 \cite{Hirata2000,Kallay2001,Kallay2004,Evangelista2006,DeYonker2012,DeYonker2014}
with basis sets that have not yet gone beyond the cc-pVDZ-DK basis
set \cite{DeYonker2014}). Furthermore, 3e$^{-}$ is the limit at
which the integrals have been expressed analytically for explicitly
correlated Slater wavefunctions, so treating 6e$^{-}$ would either
require numerically calculating the integrals (which would be too
slow even for small basis sets), or explicitly correlated Gaussian
wavefunctions (which do not necessarily have the correct short- and
long-range behavior). Therefore, beyond a certain distance the analytic
expressions ignoring wavefunction overlap but using long-range constants
for Li based on 3e$^{-}$ }\textcolor{black}{\emph{ab initio }}\textcolor{black}{calculations,
are expected to be more accurate than the 6e$^{-}$}\textcolor{black}{\emph{
ab initio}}\textcolor{black}{{} calculations of \cite{Musia2014} that
include wavefunction overlap. The distance at which this trade-off
begins to lean in favor of the analytic expressions is heuristically
given by the Le~Roy radius \cite{LeRoy1973,LeRoy1974,Ji1995}. }

\begin{figure*}
\protect\caption{Point-wise original, and analytic MLR potentials for $3b(3^{3}\Pi_{u})$
representing the \emph{ab initio} calculations of \cite{Musia2014}.
The inset shows the long-range behavior in Le~Roy space: it demonstrates
that the original\emph{ ab initio} points unphysically dip below the
theoretical curve, while the MLR behaves correctly. \label{fig:3b}}

\includegraphics[width=1\textwidth]{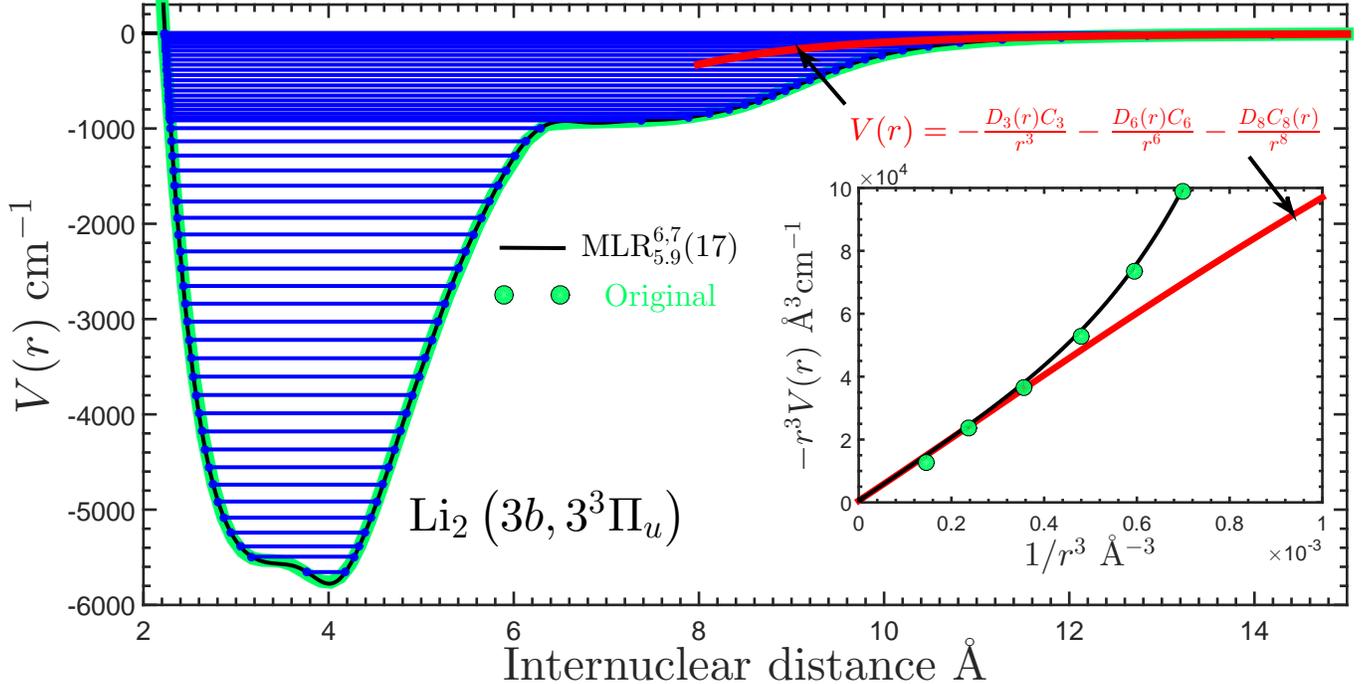}
\end{figure*}

\textcolor{black}{Another advantage of using the analytic expressions,
is that the }\textcolor{black}{\emph{ab initio}}\textcolor{black}{{}
calculations of \cite{Musia2014} do not include the effect of spin-orbit
coupling, but for alkali atoms dissociating to $nS+n^{\prime}P$ asymptotes,
the effect of spin-orbit coupling at long-range has been determined
analytically \cite{Bussery1985,Aubert-Frecon1998}. Although all papers
discussing these analytic expressions to date only mention $nS+nP$
asymptotes, the expressions are also the same for $nS+n^{\prime}P$
asymptotes when $n\ne n^{\prime}$ \cite{Aubert-Frecon2015}. }

\subsection{Le\,Roy radii}

The $m$-dependent Le~Roy radius is given by Ji \emph{et al.} \cite{Ji1995}:

\begin{equation}
R_{{\rm LR}-m}\equiv2\sqrt{3}\left(\langle nlm|z^{2}|nlm\rangle^{\nicefrac{1}{2}}+\langle n^{\prime}l^{\prime}m^{\prime}|z^{2}|n^{\prime}l^{\prime}m^{\prime}\rangle^{\nicefrac{1}{2}}\right),\label{eq:mLeRoyRadius}
\end{equation}
where for hydrogen-like atoms we have \cite{Ji1995}:

\begin{equation}
\langle nlm|z^{2}|nlm\rangle^{\nicefrac{1}{2}}=\left(\frac{1}{3}-\frac{2}{3}\frac{3m^{2}-l(l+1)}{(2l+3)(2l-1)}\right)^{\nicefrac{1}{2}}\langle nl|r^{2}|nl\rangle^{\nicefrac{1}{2}}
\end{equation}
and for $l=0$ we have (because $m$ is also $0$) \cite{Ji1995}:
\begin{equation}
\langle nlm|z^{2}|nlm\rangle^{\nicefrac{1}{2}}=\frac{1}{\sqrt{3}}\langle r^{2}\rangle^{\nicefrac{1}{2}}.
\end{equation}
This means that if both atoms of a diatomic molecule are in $S$ states,
the fact that $l=l^{\prime}=0$ reduces Eq. \ref{eq:mLeRoyRadius}
to the original Le~Roy radius of \cite{LeRoy1973,LeRoy1974}:

\begin{eqnarray}
r_{{\rm LR}} & \equiv & 2\left(\langle r_{{\rm A}}^{2}\rangle^{\nicefrac{1}{2}}+\langle r_{{\rm B}}^{2}\rangle^{\nicefrac{1}{2}}\right).
\end{eqnarray}

However for a hydrogenic atom with $l\ne0$ we have \cite{Ji1995}:

\begin{equation}
\langle nl|r^{2}|nl\rangle^{\nicefrac{1}{2}}=a_{\mu}\frac{n^{2}}{Z}\left(1+\frac{3}{2}\left(1-\frac{l(l+1)-\nicefrac{1}{3}}{n^{2}}\right)\right)^{\nicefrac{1}{2}},
\end{equation}
where $Z$ is the effective nuclear charge, $a_{\mu}=a_{0}\frac{m_{N}}{\mu}$
is the Bohr radius scaled by the ratio of the mass of the nucleus
$m_{N}$ to the reduced mass $\mu$ of the atom.  And for alkali atoms,
the principal quantum number $n$ is replaced by $n-\alpha(l)$, where
$\alpha(l)$ is the quantum defect and can be found in standard references
such as Ref {[}29{]} of \cite{Ji1995}. Using $\alpha(p)=1.59$ for
Li, assuming that $\mu_{^{6}{\rm Li}}=\frac{m_{e}M_{^{6}{\rm Li}}}{m_{e}+M_{^{6}{\rm Li}}}\approx m_{e}=1\,\mbox{a.u.}$,
and using $\langle r^{2}\rangle$ values calculated in \cite{Zhang2007},
we are able to calculate Eq. \ref{eq:mLeRoyRadius} for Li$_{2}$
molecules dissociating to various asymptotes, and we present these
in Table \ref{tab:leroyRadii}. 

\begin{table}
\protect\caption{$m$-dependent Le~Roy radii for Li$_{2}$ electronic states that
approach $2s+ml$ in Hartree atomic units and spectroscopic units,
and the constituent quantities that are used to calculate these radii.
\label{tab:leroyRadii}}

\centering{}%
\begin{tabular}{cr@{\extracolsep{0pt}.}lr@{\extracolsep{0pt}.}lr@{\extracolsep{0pt}.}lr@{\extracolsep{0pt}.}l}
\noalign{\vskip2mm}
 & \multicolumn{2}{c}{{\scriptsize{}$\langle r^{2}\rangle$ }} & \multicolumn{2}{c}{{\scriptsize{}$\langle nlm|z^{2}|nlm\rangle^{\nicefrac{1}{2}}$ }} & \multicolumn{2}{c}{{\scriptsize{}$r_{{\rm LR-m}}(2s+ml)$ }} & \multicolumn{2}{c}{{\scriptsize{}$r_{{\rm LR-m}}(2s+ml)$ }}\tabularnewline
 & ~~~~~a&u. & ~~~~~a&u. & ~~~~~a&u. & \multicolumn{2}{c}{~~~~~\AA{}}\tabularnewline[2mm]
\hline 
\noalign{\vskip2mm}
$2s$ & 17&47 \cite{Zhang2007} & 17&47 \cite{Zhang2007} & 16&7189  & 8&8473\tabularnewline
$2p$ & 27&06 \cite{Zhang2007} & 0&3810 & 27&0 & 14&0\tabularnewline
$3p$ & 168&69 \cite{Zhang2007} & 0&5910 & 40&0 & 20&1\tabularnewline[2mm]
\hline 
\end{tabular}
\end{table}

\begin{figure*}
\protect\caption{Point-wise original, and analytic MLR potentials for $3B(3^{1}\Pi_{1_{u}})$
representing the \emph{ab initio} calculations of \cite{Musia2014}.
The inset shows the long-range behavior in Le~Roy space.\label{fig:3B}}

\includegraphics[width=1\textwidth]{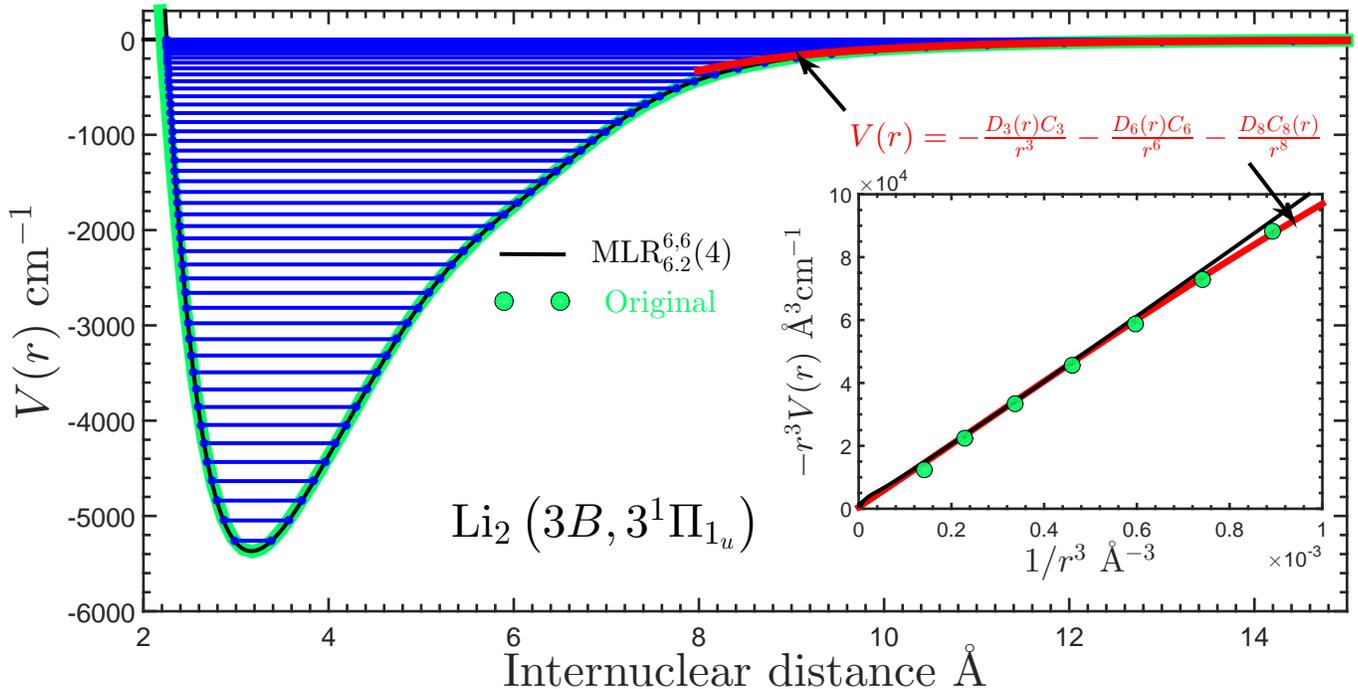}
\end{figure*}

\subsection{Long-range theory\label{sub:Long-range-theory}}

It is well-known that for large internuclear distances, the MLR model
becomes, \cite{Dattani2011}:

\begin{equation}
V(r)\simeq\mathfrak{D}_{e}-u(r)+\cdots.\label{eq:MLRlongRangeWithoutQuadraticTerm}
\end{equation}

Therefore, we can define $u(r)$ to be the analytic expression describing
the theoretical interaction between the constituent atoms of the molecule.
Each $^{2S+1}\Pi_{u/g}$ state considered in this paper has $\Omega_{u/g}$
daughter states resulting from the spin-orbit coupling that lifts
the degeneracy. For the $2_{u}$ and $2_{g}$ states, which are daughters
of $3b(3^{3}\Pi_{g})$ and $3d(3^{3}\Pi_{g})$ respectively, no other
$\Omega=2$ state with the same $u/g$ symmetry approaches the $2S+3P$
asymptote, so the potential energy curves at long-range are not strongly
influenced by other electronic states. Therefore, these states have
the simplest form for $u(r)$:

\begin{widetext}
\begin{eqnarray}
u^{2_{\nicefrac{u}{g}}}(r) & = & -\left(\Delta E-\sum_{\substack{m=3,6,8,\\
9,10,11,\ldots
}
}\frac{C_{m}^{\left(^{3}\Pi_{\nicefrac{u}{g}}\right)}}{r^{m}}\right)\\
 & = & -\left(\Delta E-\frac{C_{3}^{\left(^{3}\Pi_{\nicefrac{u}{g}}\right)}}{r^{3}}-\frac{C_{6}^{\left(^{3}\Pi_{\nicefrac{u}{g}}\right)}}{r^{6}}-\frac{C_{8}^{\left(^{3}\Pi_{\nicefrac{u}{g}}\right)}}{r^{8}}-\frac{C_{9}^{\left(^{3}\Pi_{\nicefrac{u}{g}}\right)}}{r^{9}}-\frac{C_{10}^{\left(^{3}\Pi_{\nicefrac{u}{g}}\right)}}{r^{10}}-\frac{C_{11}^{\left(^{3}\Pi_{\nicefrac{u}{g}}\right)}}{r^{11}}\cdots\right),\label{eq:2u/g}
\end{eqnarray}
where the zero of energy is the Li$(2S_{\nicefrac{1}{2}})$ + Li($3P_{\nicefrac{1}{2}}$)
asymptote, and $\Delta E\equiv\Delta E_{3^{2}P_{3/2}\leftarrow3^{2}P_{1/2}}$
is included since the $2_{u}$ and $2_{g}$ states both dissociate
to Li$(2S_{\nicefrac{1}{2}})$ + Li($3P_{\nicefrac{3}{2}}$). The
$3b(3^{3}\Pi_{u})$ state additionally has a daughter state of $1_{u}$
symmetry, along with $3B(3^{1}\Pi_{u})$; and the $3d(3^{3}\Pi_{g})$
state additionally has a daughter state of $1_{g}$ symmetry, along
with $3C(3^{1}\Pi_{g})$. Since all $1_{\nicefrac{u}{g}}$ states
approaching $2S+3P$ have two other $1_{u/g}$ states of the same
$\nicefrac{u}{g}$ symmetry approaching $2S+3P$, the $u(r)$ for
these states is defined as the highest, middle, or lowest energy eigenvalue
of the following $3\times3$ matrix (including the prefactor of $-1$)
depending on whether the state in question is the lowest, middle,
or highest in energy respectively: 

\begin{equation}
\mathbf{u}^{1_{\nicefrac{u}{g}}}(r)=
\end{equation}

{\tiny{}
\begin{equation}
\hspace{-5mm}-\left(\begin{array}{ccc}
\frac{1}{3}{\displaystyle \sum_{m}\frac{C_{m}^{\left(^{\nicefrac{3}{1}}\Sigma_{\nicefrac{u}{g}}^{+}\right)}+C_{m}^{\left(^{\nicefrac{1}{3}}\Pi_{\nicefrac{u}{g}}\right)}+C_{m}^{\left(^{\nicefrac{3}{1}}\Pi_{\nicefrac{u}{g}}\right)}}{r^{m}}} & \frac{1}{3\sqrt{2}}{\displaystyle \sum_{m}\frac{-2C_{m}^{\left(^{\nicefrac{3}{1}}\Sigma_{\nicefrac{u}{g}}^{+}\right)}+C_{m}^{\left(^{\nicefrac{1}{3}}\Pi_{\nicefrac{u}{g}}\right)}+C_{m}^{\left(^{\nicefrac{3}{1}}\Pi_{\nicefrac{u}{g}}\right)}}{r^{m}}} & \frac{1}{\sqrt{6}}{\displaystyle \sum_{m}\frac{-C_{m}^{\left(^{\nicefrac{1}{3}}\Pi_{\nicefrac{u}{g}}\right)}+C_{m}^{\left(^{\nicefrac{3}{1}}\Pi_{\nicefrac{u}{g}}\right)}}{r^{m}}}\\
\frac{1}{3\sqrt{2}}{\displaystyle \sum_{m}\frac{-2C_{m}^{\left(^{\nicefrac{3}{1}}\Sigma_{\nicefrac{u}{g}}^{+}\right)}+C_{m}^{\left(^{\nicefrac{1}{3}}\Pi_{\nicefrac{u}{g}}\right)}+C_{m}^{\left(^{\nicefrac{3}{1}}\Pi_{\nicefrac{u}{g}}\right)}}{r^{m}}} & \Delta E+\frac{1}{6}{\displaystyle \sum_{m}\frac{4C_{m}^{\left(^{\nicefrac{3}{1}}\Sigma_{\nicefrac{u}{g}}^{+}\right)}+C_{m}^{\left(^{\nicefrac{1}{3}}\Pi_{\nicefrac{u}{g}}\right)}+C_{m}^{\left(^{\nicefrac{3}{1}}\Pi_{\nicefrac{u}{g}}\right)}}{r^{m}}} & \frac{1}{2\sqrt{3}}{\displaystyle \sum_{m}\frac{-C_{m}^{\left(^{\nicefrac{1}{3}}\Pi_{\nicefrac{u}{g}}\right)}+C_{m}^{\left(^{\nicefrac{3}{1}}\Pi_{\nicefrac{u}{g}}\right)}}{r^{m}}}\\
\frac{1}{\sqrt{6}}{\displaystyle \sum_{m}\frac{-C_{m}^{\left(^{\nicefrac{1}{3}}\Pi_{\nicefrac{u}{g}}\right)}+C_{m}^{\left(^{\nicefrac{3}{1}}\Pi_{\nicefrac{u}{g}}\right)}}{r^{m}}} & \frac{1}{2\sqrt{3}}{\displaystyle \sum_{m}\frac{-C_{m}^{\left(^{\nicefrac{1}{3}}\Pi_{\nicefrac{u}{g}}\right)}+C_{m}^{\left(^{\nicefrac{3}{1}}\Pi_{\nicefrac{u}{g}}\right)}}{r^{m}}} & -\Delta E+\frac{1}{2}{\displaystyle \sum_{m}\frac{C_{m}^{\left(^{\nicefrac{1}{3}}\Pi_{\nicefrac{u}{g}}\right)}+C_{m}^{\left(^{\nicefrac{3}{1}}\Pi_{\nicefrac{u}{g}}\right)}}{r^{m}}}
\end{array}\right).\label{eq:1u/g3x3}
\end{equation}
}{\tiny \par}

\begin{figure*}
\protect\caption{Point-wise original, and analytic MLR potentials for $3C(3^{1}\Pi_{g})$
representing the \emph{ab initio} calculations of \cite{Musia2014}.
The top inset shows the long-range behavior in Le~Roy space, and
the bottom inset shows that the MLR successfully captures the tiny
second minimum which has a depth of $\approx8.5$~cm$^{-1}$.\label{fig:3C}}

\includegraphics[width=1\textwidth]{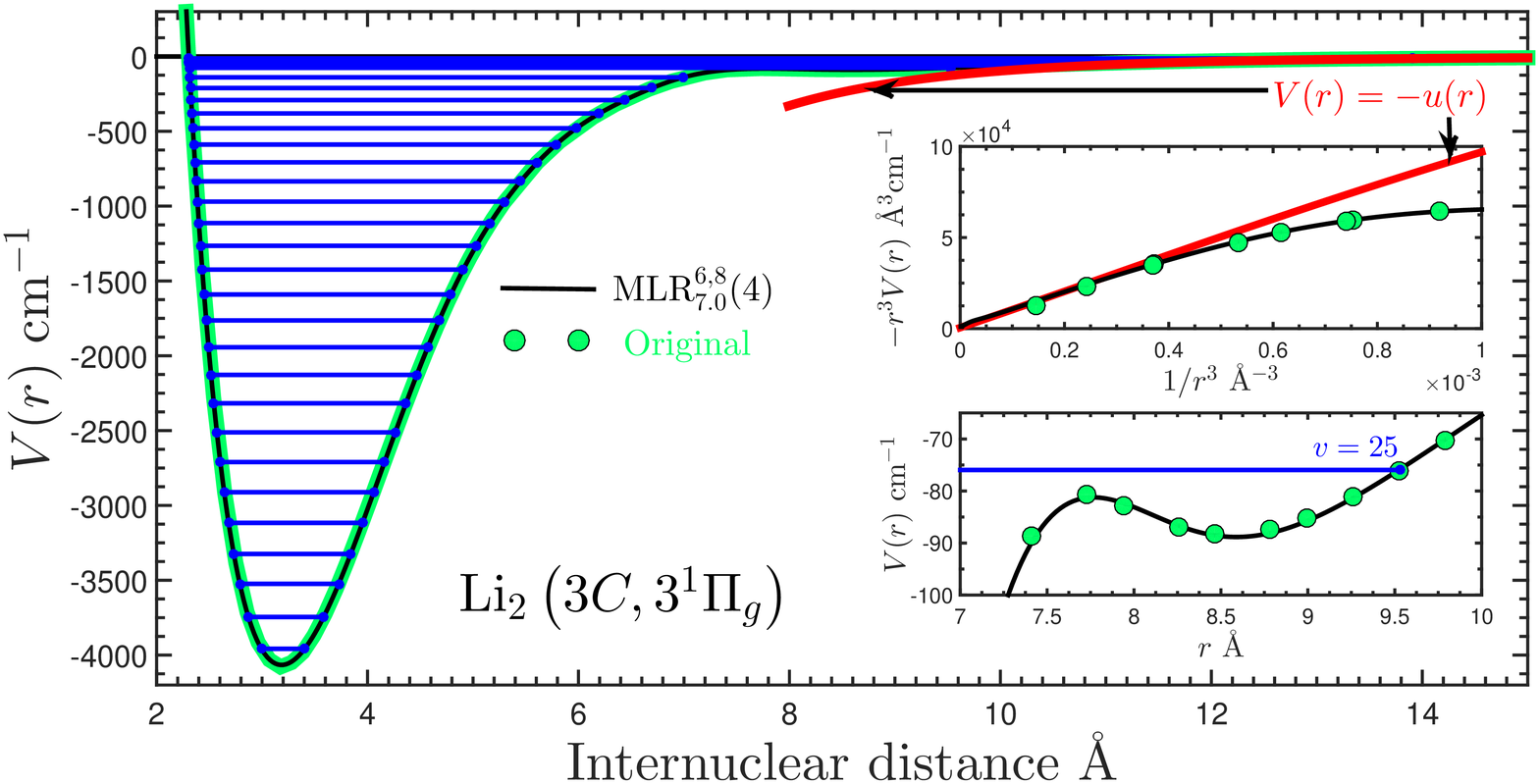}
\end{figure*}
The notation $^{\nicefrac{1}{3}}\Lambda_{u/g}$ means $(^{1}\Lambda_{u}$
or $^{3}\Lambda_{g})$. The zero of energy is once again the Li$(2S_{\nicefrac{1}{2}})$
+ Li($3P_{\nicefrac{1}{2}}$) asymptote. Finally, the $3b$ state
additionally has $0_{u}^{+}$ and $0_{u}^{-}$ daughter states, and
the $3d$ state additionally has $0_{g}^{+}$ and $0_{g}^{-}$ daughter
states. Since all $0_{\nicefrac{u}{g}}^{\nicefrac{+}{-}}$ states
approaching $2S+3P$ have one other $0_{\nicefrac{u}{g}}^{\nicefrac{+}{-}}$
state of the same $\nicefrac{u}{g}$ symmetry and the same $\nicefrac{+}{-}$
symmetry approaching $2S+3P$, the $u(r)$ for these states is defined
as the higher, or lower energy eigenvalue of the following $2\times2$
matrix (including the prefactor of $-1$) depending on whether the
state in question is lower, or higher in energy respectively:

\begin{eqnarray}
\mathbf{u}^{0_{\nicefrac{u}{g}}^{\nicefrac{+}{-}}}(r) & =- & \begin{pmatrix}\frac{1}{3}{\displaystyle \sum_{\substack{m=3,6,8\\
9,10,11,\ldots
}
}}\frac{C_{m}^{\left(^{\nicefrac{3}{1}}\Sigma_{\nicefrac{u}{g}}^{+}\right)}+2C_{m}^{\left(^{\nicefrac{1}{3}}\Pi_{\nicefrac{u}{g}}\right)}}{r^{m}} & \quad\;\frac{\sqrt{2}}{3}{\displaystyle \sum_{\substack{m=3,6,8\\
9,10,11,\ldots
}
}}\frac{C_{m}^{\left(^{\nicefrac{3}{1}}\Sigma_{\nicefrac{u}{g}}^{+}\right)}-C_{m}^{\left(^{\nicefrac{1}{3}}\Pi_{\nicefrac{u}{g}}\right)}}{r^{3}}\\
\frac{\sqrt{2}}{3}{\displaystyle \sum_{\substack{m=3,6,8\\
9,10,11,\ldots
}
}}\frac{C_{m}^{\left(^{\nicefrac{3}{1}}\Sigma_{\nicefrac{u}{g}}^{+}\right)}-C_{m}^{\left(^{\nicefrac{1}{3}}\Pi_{\nicefrac{u}{g}}\right)}}{r^{m}} & \qquad\quad-\Delta E+\frac{2}{3}{\displaystyle \sum_{\substack{m=3,6,8\\
9,10,11,\ldots
}
}}\frac{C_{m}^{\left(^{\nicefrac{3}{1}}\Sigma_{\nicefrac{u}{g}}^{+}\right)}+C_{m}^{\left(^{\nicefrac{1}{3}}\Pi_{\nicefrac{u}{g}}\right)}}{r^{m}}
\end{pmatrix}.\label{eq:0-}
\end{eqnarray}
The zero of energy is once again the Li$(2S_{\nicefrac{1}{2}})$ +
Li($3P_{\nicefrac{1}{2}}$) asymptote.

\end{widetext}

\begin{figure*}
\protect\caption{Point-wise original, and analytic MLR potentials for $3d(3^{3}\Pi_{g})$
representing the \emph{ab initio} calculations of \cite{Musia2014}.
The top inset shows the long-range behavior in Le~Roy space, and
the bottom inset shows that the MLR successfully captures the tiny
second minimum which has a depth of $\approx13.5$~cm$^{-1}$.\label{fig:3d}}

\includegraphics[width=1\textwidth]{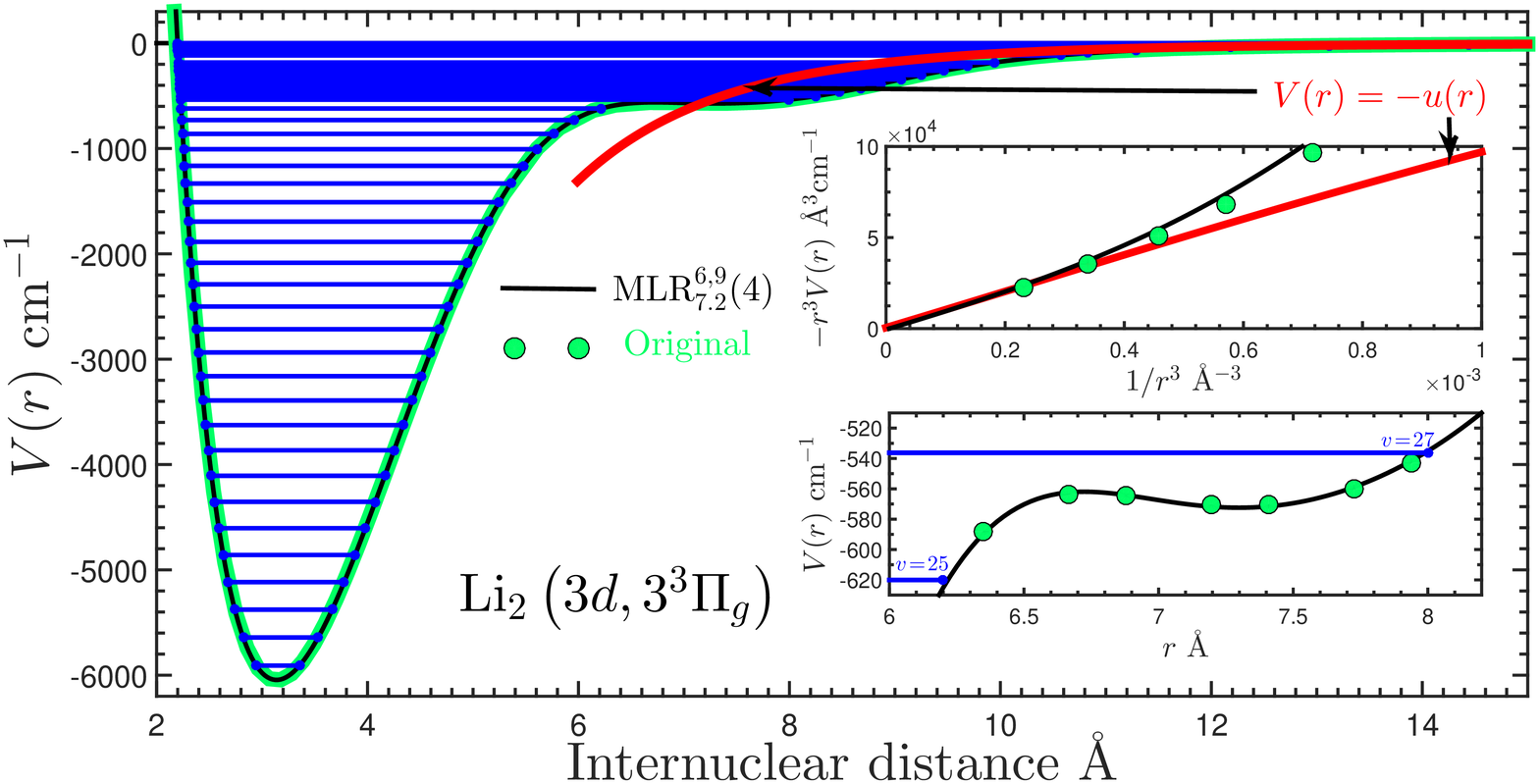}
\end{figure*}

Since the leading term not shown in Eq. \ref{eq:MLRlongRangeWithoutQuadraticTerm}
is $\frac{u(r)^{2}}{4\mathfrak{D}_{e}}$, the contribution of the
$C_{3}$ terms to the long-range form of the potential, will interfere
with the desired $C_{6}$ and $C_{8}$ terms, and all $C_{9}$ and
$C_{11}$ terms will therefore have spurious contributions from the
cross-terms formed by the products of the $C_{3}$ terms with the
$C_{6}$ and $C_{8}$ terms respectively. We fix this in the same
way as was done for $C_{6}$ and $C_{9}$ in \cite{Dattani2008,LeRoy2009,Dattani2011,Semczuk2013,Gunton2013,Dattani2015a},
 by applying a transformation to all $C_{6}$, $C_{9}$, and this
time also $C_{11}$ terms:

\begin{eqnarray}
C_{6} & \rightarrow & C_{6}+\frac{C_{3}^{2}}{4\mathfrak{D}_{e}}\label{eq:C6dattaniCorrction}\\
C_{9} & \rightarrow & C_{9}+\frac{C_{3}C_{6}}{2\mathfrak{D}_{e}},\label{eq:C9dattaniCorrection}\\
C_{11} & \rightarrow & C_{11}+\frac{C_{3}C_{8}}{2\mathfrak{D}_{e}}.
\end{eqnarray}

{\scriptsize{}}{\scriptsize \par}

\noindent {\scriptsize{}}where the transformation in Eq. \ref{eq:C6dattaniCorrction}
has to be made first due to Eq. \ref{eq:C9dattaniCorrection}'s dependence
on $C_{6}$.

Additionally, the long-range formulas in terms of $C_{m}$ constants
in Eqs. \ref{eq:2u/g},\ref{eq:1u/g3x3},\ref{eq:0-} were derived
under the assumption that two free atoms are interacting with each
other, and there is no overlap of the electrons' wavefunctions as
would be in a bound molecule. To take into account the effect of electron
overlap, we use the damping function form from \cite{LeRoy2011}:

\begin{eqnarray}
C_{m} & \rightarrow & C_{m}D_{m}^{(s)}\left(r\right)\\
D_{m}^{(s)}\left(r\right) & \equiv & \left(1-e^{-\left(\frac{b^{(s)}\rho r}{m}+\frac{c^{(s)}\left(\rho r\right)^{2}}{\sqrt{m}}\right)}\right)^{m+s},\label{eq:damping (last MLR definition)}
\end{eqnarray}

\noindent where for interacting atoms A and B, $\rho\equiv\rho_{{\rm AB}}=\frac{2\rho_{{\rm A}}\rho_{{\rm B}}}{\rho_{{\rm A}}\rho_{{\rm B}}},$
in which $\rho_{{\rm X}}\equiv\left(\nicefrac{I^{{\rm X}}}{I^{{\rm H}}}\right)^{\nicefrac{2}{3}}$
is defined in terms of the ionization potentials of atom X, denoted
$\left(I^{{\rm X}}\right)$, and hydrogen $\left(I^{{\rm H}}\right)$.
We use $s=-1$, which as shown in \cite{LeRoy2011}, means that the
MLR potential has the physically desired behavior $V(r)\simeq\nicefrac{1}{r^{2}}$
in the limit as $r\rightarrow0$. For $s=-1$, the system independent
parameters take the values $b^{(-1)}=3.30$, and $c^{(-1)}=0.423$
\cite{LeRoy2011}.

\subsection{Long-range constants\label{sub:Long-range-constants}}

For electronic states of Li$_{2}$ approaching the $2S+2P$ asymptote,
the $C_{3,6,8}$ constants for all electronic state symmetries have
been calculated with finite-mass corrections for $^{6}$Li and $^{7}$Li
\cite{Tang2009}, and even an attempt at relativistic corrections
has been made for the $C_{3}$ constants \cite{Tang2015,Tang2010a}.
Furthermore, for $2S+2P$, third-order perturbation theory has been
used to calculate non-relativistic infinite-mass values for $C_{9}$
and $C_{11}$ \cite{Tang2011} , meaning that it was possible to also
include the non-relativistic infinite-mass value of $C_{10}$ calculated
in \cite{Zhang2007}.

The situation is much less convenient for $2S+3P$. No third-order
perturbation theory calculation has been done for $C_{9}$ or $C_{11}$,
and without $C_{9}$ it does not make sense to include the $C_{10}$
value, which was calculated in the same study as for the $2S+2P$
asymptote \cite{Zhang2007}. Also, no finite-mass or relativistic
corrections have been calculated for the $C_{3,6,8}$ values associated
with $2S+3P$. Nevertheless, we have available the non-relativistic
infinite-mass values for $C_{3,6,8}$ that were calculated in \cite{Zhang2007},
and these were reported with an order of magnitude higher precision
than in the very highly cited 1995 paper of Marinescu and Dalgarno
\cite{Marinescu1995}, and only one order of magnitude lower precision
than the $2S+2P$ values which are known (see Table 2 in \cite{Dattani2015a}
for a list of the best known $C_{m}$ constants for each symmetry
approaching $2S+2P$). All $C_{m}$ constants that are used in this
study for $2S+3P$ are given in Table \ref{tab:longRangeConstants}.

\begin{table}
\begin{raggedright}
\protect\caption{The best currently available long-range constants for Li$_{2}$ electronic
states that dissociate to $2S+3P$ (in Hartree atomic units). All
values come from \cite{Zhang2007} and were calculated without relativistic
corrections, and under the assumption that both Li nuclei have infinite
mass. \label{tab:longRangeConstants}}

\par\end{raggedright}

\begin{tabular*}{1\columnwidth}{@{\extracolsep{\fill}}cr@{\extracolsep{0pt}.}lr@{\extracolsep{0pt}.}lr@{\extracolsep{0pt}.}lc}
\hline 
\noalign{\vskip2mm}
\multirow{1}{*}{} & \multicolumn{2}{c}{$^{\nicefrac{1}{3}}\Sigma_{\nicefrac{u}{g}}$} & \multicolumn{2}{c}{$^{\nicefrac{3}{1}}\Sigma_{\nicefrac{u}{g}}$} & \multicolumn{2}{c}{$^{\nicefrac{1}{3}}\Pi_{\nicefrac{u}{g}}$} & \multicolumn{1}{c}{$^{\nicefrac{3}{1}}\Pi_{\nicefrac{u}{g}}$}\tabularnewline[2mm]
\hline 
\noalign{\vskip2mm}
$C_{3}$ & 0&0033314  & $-$0&0033314 & $-$0&0016657 & 0.0016657\tabularnewline
$C_{6}$ & 3&8236$\times10^{4}$ & 3&8236$\times10^{4}$ & 2&0282$\times10^{4}$ & 2.0282$\times10^{4}$\tabularnewline
$C_{8}$ & 2&4870$\times10^{7}$ & 2&3183$\times10^{7}$ & 7&8976$\times10^{5}$ & 3.7222$\times10^{5}$\tabularnewline[2mm]
\hline 
\end{tabular*}

\rule[0.1ex]{1\columnwidth}{0.5pt}
\end{table}

\begin{table*}
\protect\caption{Parameters defining the MLR$_{p,q}^{r_{{\rm ref}}}(N_{\beta})$ potentials
fitted to \emph{ab initio} points from \cite{Musia2014}, and with
long-range functions $u(r)$ defined according to the descriptions
in section \ref{sub:Long-range-constants}, and with long-range constants
presented in Table \ref{tab:longRangeConstants}. Numbers in parentheses
are 95\% confidence limit uncertainties in the last digit(s) shown,
calculated from the least-squares fitting procedure.\label{tab:MLRpotentials}}

\begin{tabular*}{1\textwidth}{@{\extracolsep{\fill}}r@{\extracolsep{0pt}.}l>{\centering}p{0.17\textwidth}r@{\extracolsep{0pt}.}l>{\centering}p{0.17\textwidth}r@{\extracolsep{0pt}.}l>{\centering}p{0.17\textwidth}r@{\extracolsep{0pt}.}l>{\centering}p{0.17\textwidth}r@{\extracolsep{0pt}.}lr@{\extracolsep{0pt}.}l}
\hline 
\noalign{\vskip2mm}
\multicolumn{6}{>{\centering}p{0.33\textwidth}}{$3b(3^{3}\Pi_{u})$} & \multicolumn{3}{c}{$3B(3^{1}\Pi_{u})$} & \multicolumn{3}{c}{$3C(3^{1}\Pi_{g})$} & \multicolumn{4}{c}{$3d(3^{3}\Pi_{g})$}\tabularnewline
\multicolumn{6}{c}{MLR$_{6,7}^{5.9}(17)$} & \multicolumn{3}{c}{MLR$_{6,6}^{6.2}(4)$} & \multicolumn{3}{c}{MLR$_{6,8}^{7.0}(8)$} & \multicolumn{4}{c}{MLR$_{6,9}^{7.2}(4)$}\tabularnewline[2mm]
\hline 
\noalign{\vskip2mm}
\multicolumn{2}{c}{$\mathfrak{D}_{e}$} & 5765.593\textcolor{black}{~~cm$^{-1}$} & \multicolumn{2}{c}{} &  & \multicolumn{2}{c}{$\mathfrak{D}_{e}$} & 5368.8(38)\textcolor{black}{~~cm$^{-1}$} & \multicolumn{2}{c}{$\mathfrak{D}_{e}$} & \textcolor{black}{$4066.0467$~~cm$^{-1}$} & \multicolumn{2}{c}{$\mathfrak{D}_{e}$} & \multicolumn{2}{c}{\textcolor{black}{$6045.135$~~cm$^{-1}$}}\tabularnewline
\multicolumn{2}{c}{$r_{e}$} & 3.984~527(23)\textcolor{black}{~$\mbox{\AA}$} & \multicolumn{2}{c}{} &  & \multicolumn{2}{c}{$r_{e}$} & 3.165~8(18)\textcolor{black}{~$\mbox{\AA}$} & \multicolumn{2}{c}{$r_{e}$} & 3.184~402(23)\textcolor{black}{~$\mbox{\AA}$} & \multicolumn{2}{c}{$r_{e}$} & 3&137~142(23)\textcolor{black}{~$\mbox{\AA}$}\tabularnewline
\multicolumn{2}{c}{\textbf{\textcolor{black}{\small{}$\beta_{0}$}}} & 0.378369938 & \multicolumn{2}{c}{\textbf{\textcolor{black}{\small{}$\beta_{10}$}}} & -3213.8074 & \multicolumn{2}{c}{\textbf{\textcolor{black}{\small{}$\beta_{0}$}}} & -0.8509 & \multicolumn{2}{c}{\textbf{\textcolor{black}{\small{}$\beta_{0}$}}} & 0.78516117 & \multicolumn{2}{c}{\textbf{\textcolor{black}{\small{}$\beta_{0}$}}} & 0&78516117\tabularnewline
\multicolumn{2}{c}{\textbf{\textcolor{black}{\small{}$\beta_{1}$}}} & 0.918196351 & \multicolumn{2}{c}{\textbf{\textcolor{black}{\small{}$\beta_{11}$}}} & 1673.5923 & \multicolumn{2}{c}{\textbf{\textcolor{black}{\small{}$\beta_{1}$}}} & -0.592 & \multicolumn{2}{c}{\textbf{\textcolor{black}{\small{}$\beta_{1}$}}} &  2.3951022 & \multicolumn{2}{c}{\textbf{\textcolor{black}{\small{}$\beta_{1}$}}} & 2&39510218\tabularnewline
\multicolumn{2}{c}{\textbf{\textcolor{black}{\small{}$\beta_{2}$}}} & 2.08782371 & \multicolumn{2}{c}{\textbf{\textcolor{black}{\small{}$\beta_{12}$}}} & 6482.429 & \multicolumn{2}{c}{\textbf{\textcolor{black}{\small{}$\beta_{2}$}}} & -0.038 & \multicolumn{2}{c}{\textbf{\textcolor{black}{\small{}$\beta_{2}$}}} &  6.950787 & \multicolumn{2}{c}{\textbf{\textcolor{black}{\small{}$\beta_{2}$}}} & 6&9507868\tabularnewline
\multicolumn{2}{c}{\textbf{\textcolor{black}{\small{}$\beta_{3}$}}} & 0.96103485 & \multicolumn{2}{c}{\textbf{\textcolor{black}{\small{}$\beta_{13}$}}} & -1705.759 & \multicolumn{2}{c}{\textbf{\textcolor{black}{\small{}$\beta_{3}$}}} & 0.74 & \multicolumn{2}{c}{\textbf{\textcolor{black}{\small{}$\beta_{3}$}}} &  -4.68625 & \multicolumn{2}{c}{\textbf{\textcolor{black}{\small{}$\beta_{3}$}}} & -4&686246\tabularnewline
\multicolumn{2}{c}{\textbf{\textcolor{black}{\small{}$\beta_{4}$}}} & -46.0405588 & \multicolumn{2}{c}{\textbf{\textcolor{black}{\small{}$\beta_{14}$}}} & -7329.75 & \multicolumn{2}{c}{\textbf{\textcolor{black}{\small{}$\beta_{4}$}}} & 0.57 & \multicolumn{2}{c}{\textbf{\textcolor{black}{\small{}$\beta_{4}$}}} & -5.17796 & \multicolumn{2}{c}{\textbf{\textcolor{black}{\small{}$\beta_{4}$}}} & -5&17796\tabularnewline
\multicolumn{2}{c}{\textbf{\textcolor{black}{\small{}$\beta_{5}$}}} & -117.75187 & \multicolumn{2}{c}{\textbf{\textcolor{black}{\small{}$\beta_{15}$}}} & 822.19 & \multicolumn{2}{c}{} &  & \multicolumn{2}{c}{\textbf{\textcolor{black}{\small{}$\beta_{5}$}}} & 5.22029 & \multicolumn{2}{c}{} & \multicolumn{2}{c}{}\tabularnewline
\multicolumn{2}{c}{\textbf{\textcolor{black}{\small{}$\beta_{6}$}}} & 271.27645 & \multicolumn{2}{c}{\textbf{\textcolor{black}{\small{}$\beta_{16}$}}} & 4297.08 & \multicolumn{2}{c}{} &  & \multicolumn{2}{c}{\textbf{\textcolor{black}{\small{}$\beta_{6}$}}} &  7.5253 & \multicolumn{2}{c}{} & \multicolumn{2}{c}{}\tabularnewline
\multicolumn{2}{c}{\textbf{\textcolor{black}{\small{}$\beta_{7}$}}} & 880.09337 & \multicolumn{2}{c}{\textbf{\textcolor{black}{\small{}$\beta_{17}$}}} & -122.54 & \multicolumn{2}{c}{} &  & \multicolumn{2}{c}{\textbf{\textcolor{black}{\small{}$\beta_{7}$}}} &  -2.100 & \multicolumn{2}{c}{} & \multicolumn{2}{c}{}\tabularnewline[2mm]
\hline 
\end{tabular*}

\rule{1\textwidth}{0.5pt}
\end{table*}

\section{MLR potentials\label{sec:MLR-potentials}}

It has been suggested that fully analytic potentials \cite{Huang2003},
and specifically the MLR \cite{Pashov2008} may not have the flexibility
required to capture some features such as multiple minima and shelves
(see examples of these features appearing in $2S-3P$ potentials of
Li$_{2}$ in Figs \ref{fig:3b}-\ref{fig:3d}). While no attempt (as
far as I am aware) has thus far been made to use a fully analytic
potential to capture such features, an increasing number of applications
of the MLR potential after the publications of \cite{Huang2003,Pashov2008}
has made it a strong case for a ``universal'' potential form. MLR-type
potentials have successfully described spectroscopic data for many
electronic states of many diatomic molecules \cite{LeRoy2006,LeRoy2007,Salami2007,Shayesteh2007,LeRoy2009,Coxon2010,Stein2010,Piticco2010a,LeRoy2011,Ivanova2011,Dattani2011,Xie2011,Yukiya2013,Knockel2013,Semczuk2013,Wang2013,Li2013,Gunton2013,Meshkov2014,Dattani2014c,Coxon2015,Walji2015,Dattani2015,Dattani2015a,Dattani2014b}.
It has also become customary to fit \emph{ab initio} data for diatomic
\cite{Xiao2013,Xiao2013a,Kedziera2015,Teodoro2015,You2015,You2015a,You2015b}
and polyatomic \cite{Li2008,Li2010,Tritzant-Martinez2013,Wang2013,Li2013a,Ma2014}
systems to MLR models . Therefore, we use the MLR model in this study,
and the results here support the idea of the MLR model being a strong
candidate for a ``universal'' model for potential energy curves
and surfaces.

All MLR potentials were made by fitting to the \emph{ab initio }points
of Ref. \cite{Musia2014} with the program ${\tt DPotFIT}$ \cite{LeRoy2013}.
Since this is a non-linear least-squares fitting, `starting parameters'
are needed in order to allow ${\tt DPotFit}$ to achieve reasonable
fits. Starting parameters were obtained from the program ${\tt BetaFIT}$
\cite{LeRoy2013}. When fitting to \emph{ab initio} points, ${\tt DPotFIT}$
aims to minimize the dimensionless root mean square deviation:

\begin{equation}
dd\equiv\sqrt{\frac{1}{N_{{\rm data}}}\sum_{i=1}^{N_{{\rm data}}}\left(\frac{V_{{\rm MLR}}(i)-V_{{\rm \textit{ab initio}}}(i)}{u_{\textit{ab initio}}(i)}\right)^{2}},\label{eq:dd}
\end{equation}
where $V_{{\rm {\rm MLR}}}(i)$ and $V_{{\rm \textit{ab initio}}}(i)$
are the values of the respective potentials at the $i^{{\rm th}}$
internuclear distance value (the order of course does not matter)
and $N_{{\rm data}}$ is the total number of \emph{ab initio }points
to which the MLR potential is being fitted. $u_{\textit{ab initio}}(i)$
is the uncertainty in the $i^{{\rm th}}$ \emph{ab initio} point,
so that the MLR potential is likely to lie more closely to \emph{ab
initio} points at distances where the \emph{ab initio} calculation
is expected to be more reliable, and the requirement for the MLR potential
to match the \emph{ab initio} is less harsh in areas where the \emph{ab
initio }calculation is expected to be less accurate. 

It is extremely hard to determine accurate estimates on the uncertainties
for \emph{ab initio} points. The \emph{ab initio} points we are using
from \cite{Musia2014} were all calculated with the same basis set
(which the authors denoted by ANO-RCC+), therefore there is no indication
of the size of the basis set error. Furthermore, all of their calculations
were done with the same number of excitations included in their coupled
cluster method: FS-CCSD(2,0) only includes 1- and 2-electron excitations,
so it would be extremely unlikely to estimate the deviation from the
full 6-electron (FCI) limit. Perhaps even more importantly, the calculations
of \cite{Musia2014} neglected relativistic, spin-orbit, and non-Born-Oppenheimer
effects, so accurately estimating $u_{{\rm \textit{ab initio}}}(i)$
might seem impossible. 

However, in my recent benchmark paper \cite{Dattani2015a}, it was
shown that none of the vibrational energies associated with the \emph{ab
initio} potential from \cite{Musia2014} for the $b(1^{3}\Pi_{u})$
state of Li$_{2},$ deviated from the empirical potential's vibrational
energies by more than 12~cm$^{-1}$. Since that \emph{ab initio }potential
used the same basis set and method as their potentials for the electronic
states approaching $2S+3P$, I aimed to make $V_{{\rm MLR}}(i)-V_{{\rm \textit{ab initio}}}(i)$
less than 15~cm$^{-1}$ for all $i$ except at very small internuclear
distances near the $r=0$ singularity where the inner wall of the
potential rapidly increases, crosses the dissociation limit, and then
attains extremely large energy values. The exact values used for $u_{\textit{ab initio}}(i)$
that were used are presented in Tables \ref{tab:3bMLRvsAbInitio},
\ref{tab:3BMLRvsAbInitio}, \ref{tab:3CMLRvsAbInitio} and \ref{tab:3dMLRvsAbInitio}.
Furthermore, there are places in which it was desirable to make $u_{\textit{ab initio}}(i)$
smaller than 15~cm$^{-1}$. This was in places where the potentials
from \cite{Musia2014} had features such as tiny second minima, tiny
shelves, or any other type of abrupt change. The subsections (below)
for each electronic state will describe in detail the nature of these
features and how this affected the choice of $u_{\textit{ab initio}}(i)$
(once again the exact values are given in Tables \ref{tab:3bMLRvsAbInitio},
\ref{tab:3BMLRvsAbInitio}, \ref{tab:3CMLRvsAbInitio} and \ref{tab:3dMLRvsAbInitio}). 

The MLR model was fitted to the points from \cite{Musia2014} with
various manually adjusted values of the MLR parameters $(N_{{\rm \beta}},p,q,r_{{\rm ref}})$
in search for the lowest $dd$ according to Eq. \ref{eq:dd} such
that increasing $N_{{\rm \beta}}$ no longer reduced $dd$ significantly
further than the best $dd$ obtained with the previous increase in
$N_{\beta}$. Details for each electronic state are described in the
subsections below which focus on each state.

\subsection{The $3b(3^{3}\Pi_{u})$ state}

The first $b$ state of Li$_{2}$ dissociates to the $2S+2P$ asymptote.
A very detailed analysis of theory vs experiment for the first $b$
state of Li$_{2}$ was recently reported \cite{Dattani2015a}, which
summarized 14 different experiments providing new information on the
$b(1^{3}\Pi_{u})$ state since 1983 \cite{Engelke1983,Preuss1985,Rai1985,Xie1985b,Xie1985c,Xie1986,Rice1986,RiceStevenF.1986,Li1992,Linton1992,Li1996c,Weyh1996a,Russier1997,Lazarov2001a},
and mentioned several other papers that involved this state without
providing information on any new levels. Due to the spin-orbit coupling
between $b$ states of alkali dimers and their respective $A(^{1}\Sigma_{u}^{+})$
states, recent experimental and theoretical papers have studied the
lowest $b$ state of Rb$_{2}$, \cite{Salami2007,Drozdova2013,Tomza2013},
NaCs \cite{Zaharova2009}, KCs \cite{Kruzins2010}, RbCs \cite{Docenko2010},
Cs$_{2}$ \cite{Bai2011} and NaK \cite{Harker2015}.

It is thus surprising that no experiments have been reported for the
second $b$ state of Li$_{2}$, which dissociates to the $2P+2P$
asymptote. The present paper is concerned with the third $b$ state,
which dissociates to $2S+3P$. The \emph{ab initio }potential for
$3b(3^{3}\Pi_{u})$ from \cite{Musia2014} has a small shelf-like
feature before the minimum, and another much longer one closer to
dissociation (see Fig. \ref{fig:3b}). The first shelf is located
between $v=0$ and $v=1$, and it lasts from about $3.2-3.6$~$\mbox{\AA}$.
The second shelf starts after $v=27$ and lasts from about $6.4-8$~$\mbox{\AA}$.
Despite these fairly pronounced shelf features, at the resolution
of the \emph{ab initio} points (which is about $0.1\,\mbox{\AA}$),
the $3b$ state only has one minimum! 

In Section \ref{sec:MLR-potentials} it was mentioned that, with
the exception of points at very small values of $r$, the goal was
to match all \emph{ab initio} points of \cite{Musia2014} to within
$\pm$15~cm$^{-1}$ (and points at larger values of $r$ even better),
since this was about the level of accuracy found when comparing \emph{ab
initio} points \cite{Musia2014} to an empirical potential for $b(1^{3}\Pi_{u})$
in \cite{Dattani2015a}. Preliminary fits used such a weighting scheme
for the least-squares fitting, except with points comprising the two
shelves mentioned in the previous paragraph, weighted with much smaller
uncertainties. This was especially important for the second shelf,
which only spanned a range of $<100$~cm$^{-1}$: Because if the
discrepancy between the MLR and \emph{ab initio} was $-15$~cm$^{-1}$
at one point, and $+15$~cm$^{-1}$ at another point, the $30$~cm$^{-1}$
range of discrepancy would constitute a significant portion of the
range of the entire shelf itself. After these preliminary fits, it
was found that in order to get the MLR matching the original data
with the desired precision, it helped to decrease the uncertainties
on the non-shelf \emph{ab initio} points to slightly below $\pm15$~cm$^{-1}$.
The best fits were found with the weights shown in Table \ref{tab:3bMLRvsAbInitio}.
Once these final weights were chosen, fits were performed for 252
different combinations of the MLR parameters $(N_{\beta},p,q,r_{{\rm ref}})$,
with $4\le N_{\beta}\le17$, $6\le p\le8$, $2\le q\le8$, $(5\le r_{{\rm ref}}\le6.5)$~\AA ,
though not every point in the convex hull formed by these ranges was
used. For example, some fits were done with $(p,q)=(6,2)$ and some
fits were done with $(p,q)=(8,6)$ but it did not seem necessary to
do fits with $(p,q)=(8,2)$. The best fit was found with $(N_{\beta},p,q,r_{{\rm ref}})=(17,6,7,5.9\,$\AA ),
and had $dd=1.548$, while the best fit with $N_{\beta}=16$ was with
$(N_{\beta},p,q,r_{{\rm ref}})=(16,6,6,5.3\,\text{\AA})$ and had
$dd=2.790$. Apart from in the inner wall of the potential, the worst
discrepancy between the MLR and the original points for this $N_{\beta}=16$
case it was $>30$~cm$^{-1}$, while for this $N_{\beta}=17$ case
was $<11$~cm$^{-1}$ so it was quite easy to select the $N_{\beta}=17$
fit. Since this $N_{\beta}=17$ fit satisfied all of our desiderata,
$N_{\beta}=18$ fits were not explored.

The final MLR parameters for the chosen case are given in Table \ref{tab:MLRpotentials}.
The inset of Fig. \ref{fig:3b} shows the long-range behavior of the
MLR potential and the original \emph{ab initio }points of \cite{Musia2014}
in Le~Roy space, and compares them to the theoretical long-range
potential based on Eq. \ref{eq:2u/g} and the long-range constants
in Table \ref{tab:longRangeConstants}. The agreement is surprisingly
excellent, however after about 17.7~$\mbox{\AA}$ $(1/r^{3}\approx0.001\,8\,\mbox{\AA}^{-3})$
we see that the original points dip below the theoretical curve, which
should not happen because $C_{8}$ is attractive (see Section 4.3
of \cite{LeRoy2007}, for example). The fact that the MLR potential
matches the theoretical curve in this regard, is yet another advantage
of using an MLR to represent the \emph{ab initio} points. Furthermore,
while not shown in Fig. \ref{fig:3b}, the theoretical long-range
curve \emph{without }damping, matches the damped curve shown in the
figure to graphical accuracy at least in the range $14\,\mbox{\AA}\le r\le\infty$
$(0\le1/r^{3}=0.000\,36\,\mbox{\AA}^{-3})$, so this conclusion about
the original points spuriously dipping below the theoretical curve
is true whether or not long-range damping is considered.

\subsection{The $3B(3^{1}\Pi_{u})$ state}

The first $B$ state has a barrier before dissociating to the $2S+2P$
asymptote, and has therefore been the subject of many empirical studies
\cite{Hessel1979,Russier1994,Bouloufa1999,Cacciani2000,Bouloufa2001,Bouloufa2001a,Huang2003}.
It has also been used to study other states, such as in \cite{Bernheim1982a,Barakat1986,Linton1990,Kubkowska2007,Linton1992}.
The second $B$ state potential (sometimes called the ``$C^{1}\Pi_{u}$
state'' rather than the $2B$ state) dissociates to $2P+2P$ and
hugs the inside of the first minimum of the $2A(1^{1}\Sigma_{u}^{+})$
potential, and due to the perturbations between these $2B$ and $2A$
states, there have been many empirical studies of the $2B$ state
\cite{BARROW1960,Ishikawa1991,Weyh1996a,Ross1998,Kasahara2000,Kubkowska2007}. 

The present paper is concerned with third $B$ state. The $3B$ state
dissociates to $2S+3P$ and there has only been one experiment which
studied the third $B$ state (sometimes called the ``$D^{1}\Pi_{u}$
state'') \cite{BARROW1960}, which was over 55 years ago! The authors
of that work mentioned in their paper that they were not able to confidently
assign vibrational quantum numbers to their data, and therefore they
were only able to conclude that $T_{e}<34\,140$~cm$^{-1}$ and $\omega_{e}\approx250$~cm$^{-1}$.
In that study, the anharmonic values $x_{e}\omega_{e}$, equilibrium
rotational constants $B_{e}$, and dissociation energies $\mathfrak{D}_{e}$
were determined for the $2B$ state of Li$_{2}$ and for the $2B$
and $3B$ states of Na$_{2}$, but not for the $3B$ state of Li$_{2}$.

It is no surprise that none of the experiments on the $2B$ state
showed any indication of a barrier in the potential, because the leading
long-range term $(C_{5})$ is attractive \cite{Zhang2007}. However,
it is perhaps surprising that the \emph{ab initio} potential of \cite{Musia2014}
for the $3B$ state does not have a barrier (at least before the calculations
stopped at about 21~$\mbox{\AA}$), because the leading long-range
interaction term $(C_{3})$ for this state is repulsive \cite{Zhang2007}.
There is however a good theoretical explanation for the lack of barrier
in the $3B$ state. The long-range potential for $3B$ is not just
a simple sum of inverse powers ($-C_{3}/r^{3}-C_{6}/r^{6}\cdots$),
but it is the middle eigenvalue of the $3\times3$ spin-orbit interaction
matrix of Eq. \ref{eq:1u/g3x3}, which involves the $6a(6^{3}\Sigma_{u})$
and $3b(3^{3}\Pi_{u})$ states. For large internuclear distances,
the $6a$ state pushes down on the $3B$ state enough to remove the
barrier. Indeed, numerical calculations of the eigenvalues of the
$3\times3$ matrix show that the potential is attractive at all distances
(beyond the repulsive inner wall for $r\ll r_{e}$). This $3\times3$
matrix for $3B$ is the exact same as the one for the first $B$ state,
which \emph{does }have a barrier, but for the $3B$ state the $C_{3}$
is three orders of magnitude smaller than for the first $B$ state,
and the $C_{6}$ is one order of magnitude bigger than for the first
$B$ state \cite{Zhang2007}. This highlights the importance of using
the $3\times3$ coupling matrix because a simple inverse power sum
with a negative leading long-range term (such as the negative $C_{3}$
in the present case) is guaranteed to have a barrier, and there was
even a barrier when using the middle eigenvalue of the $3\times3$
coupling matrix for the $2S+2P$ values of $C_{m}$, but the specific
values of $C_{m}$ at $2S+3P$ seem to be past a bifurcation point
at which the barrier is lost. 

Since alkali parent states of $B(^{1}\Pi_{u})$ symmetry only have
one spin-orbit daughter state ($\Omega_{u/g}=1_{u})$, we do not need
to worry about defining the MLR long-range function $u(r)$ in a piece-wise
manner, and it is simply defined as the middle eigenvalue of the $3\times3$
interaction matrix of Eq. \ref{eq:1u/g3x3}. Also, since there are
no features such as barriers, multiple minima or shelves, the weighting
strategy was straightforward. Preliminary fits were done with all
points from \cite{Musia2014} being weighted with uncertainty $\pm15$~cm$^{-1}$
if $V(r)<-100\,$cm$^{-1}$, with $\pm5$~cm$^{-1}$ if $V(r)<-20$~cm$^{-1}$,
with $\pm1$~cm$^{-1}$ if $V(r)<1$~cm$^{-1}$ and $\pm0.5$~cm$^{-1}$
for the one point for which $-1$\,cm$^{-1}<V(r)<0$~cm$^{-1}$.
It was then found that most uncertainties could be even further reduced
without making the fitting too difficult, and that for very small
values of $r$ it was very difficult to achieve $\pm15$~cm$^{-1}$
agreement in the fit. In the end, the points from \cite{Musia2014}
for the smallest values of $r$ (with $V(r)>-4000\,$cm$^{-1}$) were
weighted with uncertainty $\pm100$~cm$^{-1}$, and all other points
were weighted with $\pm1$~cm$^{-1}$, except for the last point
which was weighted with $\pm0.5$~cm$^{-1}$ because $V(r)$ itself
was only $\approx-0.7$~cm$^{-1}$. These final weights are shown
in Table \ref{tab:3BMLRvsAbInitio}.

Once these final weights were chosen, fits were performed for 150
different combinations of the MLR parameters $(N_{\beta},p,q,r_{{\rm ref}})$,
with $3\le N_{\beta}\le5$, $6\le p\le7$, $2\le q\le7$, $(5\le r_{{\rm ref}}\le10)$~\AA ,
though not every point in the convex hull formed by these ranges was
used. The best fit with $N_{\beta}=4$ was with $(p,q,r_{{\rm ref}})=(6,6,6.2\,\text{\AA})$
which had $dd=4.307$ while the best fit found using $N_{\beta}=5$
was only marginally better ($dd=4.128)$ and no fits were found with
$N_{\beta}=3$ that had $dd<10$. Therefore, the choice of MLR model
for this electronic state was easy to make, and the final parameters
are listed in Table \ref{tab:MLRpotentials}.

\subsection{The $3C(3^{1}\Pi_{g})$ state}

The first $C(^{1}\Pi_{g})$ state dissociates to $2S+2P$ and was
not studied in detail until 1990 \cite{Miller1990}. This was 11 years
after the second state of $C(^{1}\Pi_{g})$ symmetry (sometimes called
the ``$G^{1}\Pi_{g}$ state'' since it was given this name in \cite{Bernheim1981a})
was studied in detail in 1979 \cite{Bernheim1979}. This $2C$ state
dissociates to $2P+2P$ and was studied again in a series of follow-up
papers by Bernheim \emph{et al.} \cite{Bernheim1981a,Bernheim1982,Bernheim1983}.
Impressively, empirical spectroscopic constants have also been reported
for all Rydberg states in the series $nd\pi^{1}\Pi_{g}$ for $n=3-15$
(!) \cite{Bernheim1982,Bernheim1983}. In the same paper, it was determined
that the $n=3$ state in this series is in fact the $2C$ state.

The third $C$ state is the subject of the present work, since it
dissociates to $2S+3P$. It only has one spin-orbit daughter state,
which has $1_{g}$ symmetry and couples to the $1_{g}$ daughters
of the $3c(3^{3}\Sigma_{g}^{+})$ and $3d(3^{3}\Pi_{g})$ states.
The potential energy of the $3C$ state's $1_{g}$ daughter is given
by the middle eigenvalue of the appropriate $3\times3$ matrix. Preliminary
fits were done with the same weighting scheme as for the $3B$ state,
except with the points surrounding the second minimum weighted more
strongly (with $\pm0.5$~cm$^{-1}$), since the depth of this well
is $<10$~cm$^{-1}$ and therefore it would not be satisfactory to
fit to these points with an agreement of only $\pm15$~cm$^{-1}$!
It was then found that many points could be weighted more strongly
without making the fitting too difficult. The weights were adjusted
to $\pm10$~cm$^{-1}$ for points at the smallest values of $r$
($V(r)>-3000$~cm$^{-1}$), to $\pm5$~cm$^{-1}$ for the rest of
the points with $V(r)<-1000$~cm$^{-1}$, to $\pm1$~cm$^{-1}$
for points with $V(r)<-1$~cm$^{-1}$, except for the points near
the second minimum, and the very last point for which $V(r)\approx-0.75$~cm$^{-1}$.
The final weights are presented in Table \ref{tab:3CMLRvsAbInitio}.

With these final weights, fits were performed for 302 different combinations
of the MLR parameters $(N_{\beta},p,q,r_{{\rm ref}})$, with $2\le N_{\beta}\le11$,
$6\le p\le11$, $(4\le r_{{\rm ref}}\le12)$~\AA , though not every
point in the convex hull formed by these ranges was used. The best
fit with $N_{\beta}=8$ was with $(p,q,r_{{\rm ref}})=(6,5,7.5\,\text{\AA})$
which had $dd=1.612$, but this potential did not capture the second
minimum very well (particularly, it approaches the barrier leading
to that minimum with the \emph{ab initio} point of $V(7.408\,\text{\AA})=-88.76$
cm$^{-1}$ being represented with a discrepancy of $>5.5$~cm$^{-1}$).
The best fit with $N_{\beta}=8$ which captured $V(7.408\,\text{\AA})$
with a discrepancy of $<2$~cm$^{-1}$ was with $(p,q,r_{{\rm ref}})=(6,8,7.0\,\text{\AA})$
which had $dd=3.536$. None of the $N_{\beta}=7$ models that reproduced
$V(7.408\,\text{\AA})$ with discrepancy $<2$~cm$^{-1}$ had an
overall $dd<4$, and while $N_{\beta}=9$ fits were found with discrepancies
for this point $<2$~cm$^{-1}$ and overall $dd$ as low as $2.593$,
there were no points beyond $r=2.6$~\AA{} for which the $N_{\beta}=8$
case with $dd=3.536$ misrepresented an original point by $>18$~cm$^{-1}$
(the highest discrepancy was $17.37$~cm$^{-1}$ at 3.175~$\text{\AA}$,
and among these $N_{\beta}=9$ cases, the lowest discrepancy for this
same point was 12.70~cm$^{-1}$). While deciding not to go beyond
$N_{\beta}=8$ was not an easy choice, there is not much reason to
believe that the calculation in \cite{Musia2014} for $V(3.175\text{}\text{\AA})$
is so precise that representing it more closely by $\approx5$~cm$^{-1}$
is worth adding an extra parameter. Here it is mentioned that while
the comparison against the empirical potential in \cite{Dattani2015a}
for the lowest $b$ state showed no discrepancy of $>12$~cm$^{-1}$,
that paper also noted the surprisingly small effect of Born-Oppenheimer
breakdown in that system, meaning that it is likely that the potentials
in \cite{Musia2014} for other electronic states (especially ones
approaching $2S-3P$, which seem to interact with each other more
than the ones approaching the $2S-2P$ state) will be accurate to
slightly less precision than $\pm$12~cm$^{-1}$. The final MLR parameters
for the chosen model are listed in Table \ref{tab:MLRpotentials}.

\subsection{The $3d(3^{3}\Pi_{g})$ state}

The first $d$ state has a potential energy curve which approaches
the $2S+2P$ asymptote, but the \emph{ab initio} calculations of \cite{Musia2014}
indicate that it has no bound states. Therefore, it is no surprise
that no bound levels have been found in experiments on this state,
though it was indeed involved in some experiments \cite{Li1998,Lazarov2001,Dai2005}.
While the prediction in \cite{Musia2014} that the first $d$ state
has no bound levels is likely to be true, it should be noted that
\emph{ab initio} predictions of this sort are not always reliable.
The 1995 \emph{ab initio} study of \cite{Poteau1995} predicted that
the $1^{1}\Sigma_{u}^{-}$ state would have no bound levels, but the
2014 calculations of \cite{Musia2014} found there to be a dissociation
energy of $\mathfrak{D}_{e}=$14~cm$^{-1}$ and an equilibrium harmonic
frequency of $\omega_{e}=10$~cm$^{-1}$, indicating the existence
of at least two bound vibrational levels! Likewise, the 2006 \emph{ab
initio} study of \cite{Jasik2006} found the $1^{3}\Delta_{u}$ state
to not have any bound levels, but the earlier 1995 study of \cite{Poteau1995}
and the 2014 study of \cite{Musia2014} both predicted $\mathfrak{D}_{e}\ge3430$~cm$^{-1}$
and $\omega_{e}\le255$~cm$^{-1}$.

The second $d$ state has been studied extensively. Spectroscopic
measurements for $2d$ were made in \cite{Xie1986,Yiannopoulou1995a,Li1996,Li1996a,Weyh1996a,Russier1997,Li2007},
and $2d$ was also used in various other experimental studies such
as \cite{Xie1985,Xie1986a,Rice1986,RiceStevenF.1986,Linton1992,Linton1999,Lazarov2001,Dai2005}.

The focus of this paper is on the third $d$ state, which is the only
$\Lambda=\Pi$ state dissociating to $2S+3P$ for which rovibrationally
resolved spectra have been measured, but only 13 lines were observed
(with $v=6,7,8,10$) \cite{Yiannopoulou1995a}. Hyperfine structure
was also studied experimentally for $3d$ in \cite{Li1996}, but it
was only for the $N=6,8$ levels of $v=8$, which had already been
studied without focus on hyperfine structure in \cite{Yiannopoulou1995a}.
Finally, the $3d$ state was involved in the experiments of \cite{Li1996a},
but the focus of that study was not the $3d$ state.

Since the $3d$ state has four spin-orbit daughter states ($0_{g}^{+},0_{g}^{-},1_{g},2_{g}$,
analogous with the $3b$ state), we treat the $2_{g}$ symmetry in
this paper, since there is only one state dissociating to $2S+3P$
with $2_{g}$ symmetry and therefore the long-range potential is a
simple sum of inverse-power terms rather than a complicated $2\times2$
or $3\times3$ interaction matrix. The fitting strategy was very similar
to what it was for the $3C$ state, except the $3d$ state seemed
to require stronger weighting of the points near the second minimum,
and weaker weighting of other points.  These final weights are shown
in Table \ref{tab:3BMLRvsAbInitio}.

Once these final weights were chosen, fits were performed for 238
different combinations of the MLR parameters $(N_{\beta},p,q,r_{{\rm ref}})$,
with $3\le N_{\beta}\le11$, $5\le p\le8$, $2\le q\le10$, $(5\le r_{{\rm ref}}\le7.5)$~\AA ,
though not every point in the convex hull formed by these ranges was
used, and it is noted that $p$ must be $\ge6$ in order to ensure
the correct long-range behavior \cite{Dattani2011}, but fits with
$p=5$ were still instructive to better understand the model dependence
for this potential. The best fit with $N_{\beta}=4$ and $p\ge6$
was with $(p,q,r_{{\rm ref}})=(6,7,7.2\,\text{\AA})$ which had $dd=4.292$,
only 0.043 higher than the best fit found with $p=5$. Fits with $N_{\beta}=5$
had $dd$ values as low as $2.972$ but the mentioned $N_{\beta}=4$
case did not misrepresent any of the original points beyond $2.434\,\text{\AA})$
by $>13$~cm$^{-1}$, so all desiderata were satisfied without resorting
to $N_{\beta}=5$. No $N_{\beta}=3$ cases had $dd<10$, so the choice
of MLR model for this electronic state was easy to make, and the final
parameters are listed in Table \ref{tab:MLRpotentials}.

\section{Conclusion}

Analytic MLR potentials were fitted to the \emph{ab initio} points
from \cite{Musia2014} and with correct long-range behavior incorporated
according to effects of spin-orbit coupling described in Eqs. \ref{eq:2u/g},
\ref{eq:1u/g3x3}, \ref{eq:0-} and the long-range constants in Table
\ref{tab:longRangeConstants}. Despite the potentials from \cite{Musia2014}
having unusual features such as multiple minima, barriers, and shelves,
which have never been described by an MLR-type model before, all of
these features were successfully captured with the MLR model. This
answers an age-old question of whether or not fully analytic potentials
can have the flexibility needed in order to capture such features.
Pashov's 2008 paper ``Pointwise and analytic potentials for diatomic
molecules. An attempt for critical comparison'' \cite{Pashov2008}
described lack of flexibility as one of the three drawbacks of analytic
potentials, and suggested that the MLR model may not be able to capture
double minima or shelf-like features. Five years earlier in 2003,
Huang and Le~Roy suggested in \cite{Huang2003} that Pashov's pointwise
approach would be the method of choice for potentials such as those
described in this paper:
\begin{quote}
``A particular strength of {[}Pashov's pointwise{]} model is the
fact that it has more local flexibility than do fully analytical potential
function forms, in that a shift of one potential point has only a
modest effect on the function outside its immediate neighborhood.
This would tend to make {[}Pashov's pointwise{]} model the method
of choice for cases where the potential has substantial local structure
or undergoes an abrupt change of character on a small fraction of
the overall interval, such as occurs near an avoided curve crossing.
In contrast, a change in one of the parameters defining a {[}fully
analytic{]} such as our DELR function will in general affect the potential
across the whole domain. This makes the parameters defining {[}fully
analytic{]} potentials very highly correlated and can give rise to
difficulty in achieving full unique convergence in a fit.''
\end{quote}
At the time when this quote was written, the MLR model did not exist
yet, and at the time of Pashov's paper \cite{Pashov2008}, only a
primitive (less flexible) form of the MLR existed, which was used
in just four simple cases of ground electronic states \cite{LeRoy2006,LeRoy2007,Salami2007,Shayesteh2007}.
It is possible that the notion that analytic potentials cannot capture
such special features may be attributed to the lack of diversity in
attempted applications at that early stage in time, and the lack of
some of the newer MLR features which were introduced in \cite{LeRoy2009}
and \cite{LeRoy2011} for increasing flexibility, and in \cite{Dattani2008}
for correcting the long-range behavior.

This paper also represents, to the best of my knowledge, the most
detailed study of analytic potentials for excited electronic molecular
states.

\section*{Acknowledgments}

I am indebted to Monique Aubert-Fr\'{e}con for taking the time to look
through her old notes to answer my request for advice on the analytic
form for the long-range potentials at the $nS-n^{\prime}P$ asymptotes
of alkali dimers. I also thank Michael Bromley of The University of
Queensland (Australia) for advice about the long-range constants at
the $nS-n^{\prime}P$ asymptote, and Robert J. Le~Roy of University
of Waterloo (Canada) for many helpful discussions. 

\clearpage{}

\begin{table}
\protect\caption{Quality of MLR$_{6,7}^{5.9}(17)$ fit to original \emph{ab initio
}energies of \cite{Musia2014} for the $3b(3^{3}\Pi_{2_{u}})$ state.
When an isotopologue listed in Table \ref{tab:3b v-levels} has at
least one vibrational energy level beyond the range of \emph{ab initio
}data available, approximate distances are given for the outer classical
turning points of the corresponding vibrational wavefunctions. \label{tab:3bMLRvsAbInitio}}

\begin{tabular}{r@{\extracolsep{0pt}.}lr@{\extracolsep{0pt}.}lcr@{\extracolsep{0pt}.}lr@{\extracolsep{0pt}.}l}
\hline 
\noalign{\vskip2mm}
\multicolumn{2}{c}{\textcolor{blue}{\small{}~$r$}\textcolor{blue}{~}} & \multicolumn{2}{c}{\textcolor{blue}{\small{}Original }} & \textcolor{blue}{\small{}Weight } & \multicolumn{2}{c}{\textcolor{blue}{\small{}MLR - Original }} & \multicolumn{2}{c}{\textcolor{blue}{\small{}MLR}}\tabularnewline
\multicolumn{2}{c}{\textcolor{blue}{{[}$\mbox{\AA}${]}}} & \multicolumn{2}{c}{\textcolor{blue}{\small{}{[}cm$^{-1}${]}}} & \textcolor{blue}{\small{}{[}cm$^{-1}${]}} & \multicolumn{2}{c}{\textcolor{blue}{\small{}{[}cm$^{-1}${]}}} & \multicolumn{2}{c}{\textcolor{blue}{\small{}{[}cm$^{-1}${]}}}\tabularnewline[2mm]
\hline 
\noalign{\vskip2mm}
2&328~380 & -1~501&30 & 40.00 & 195&839 & -1305&461\tabularnewline[-0.65mm]
2&434~215 & -2~661&44 & 40.00 & 9&542 & -2651&898\tabularnewline[-0.65mm]
2&540~051 & -3~556&46 & 40.00 & -79&794 & -3636&254\tabularnewline[-0.65mm]
2&645~886 & -4~232&00 & 40.00 & -104&149 & -4336&149\tabularnewline[-0.65mm]
2&751~721 & -4~727&14 & 40.00 & -89&498 & -4816&638\tabularnewline[-0.65mm]
2&857~557 & -5~075&00 & 20.00 & -57&523 & -5132&523\tabularnewline
\hline 
\end{tabular}
\end{table}

\begin{table}
\protect\caption{Vibrational energies in cm$^{-1}$ for the $3b(3^{3}\Pi_{2_{u}})$
state predicted by the MLR$_{6,7}^{5.9}(17)$ from Table \ref{tab:MLRpotentials}.\label{tab:3b v-levels}}

\begin{tabular}{r@{\extracolsep{0pt}.}lr@{\extracolsep{0pt}.}lcr@{\extracolsep{0pt}.}lr@{\extracolsep{0pt}.}l}
\hline 
\noalign{\vskip2mm}
\multicolumn{2}{c}{\textcolor{black}{\small{}~$v$}} & \multicolumn{2}{c}{\textcolor{blue}{\small{}$^{6,6}$Li$_{2}$}} & \textcolor{blue}{\small{}$^{6,7}$Li$_{2}$} & \multicolumn{2}{c}{\textcolor{blue}{\small{}$^{7,7}$Li$_{2}$}} & \multicolumn{2}{c}{\textcolor{blue}{\small{}$^{11,11}$Li$_{2}$}}\tabularnewline[2mm]
\hline 
\noalign{\vskip2mm}
\multicolumn{2}{c}{\textbf{\small{}$\mathfrak{D}_{e}$}} & \multicolumn{2}{c}{\textcolor{black}{$5\,765.593$}} & \textcolor{black}{$5\,765.593$} & \multicolumn{2}{c}{\textcolor{black}{$5\,765.593$}} & \multicolumn{2}{c}{\textcolor{black}{$5\,765.593$}}\tabularnewline[-0.65mm]
\multicolumn{2}{c}{\textbf{\textcolor{black}{\small{}$0$}}} & \multicolumn{2}{c}{$-5\,654.932$} &  -5658.488 &  -5662&222 & -5661&726\tabularnewline[-0.65mm]
\multicolumn{2}{c}{\textbf{\textcolor{black}{\small{}$1$}}} & \multicolumn{2}{c}{$-5\,493.765$} &  -5500.003 &  -5506&570 & -5505&696\tabularnewline[-0.65mm]
\multicolumn{2}{c}{\textbf{\textcolor{black}{\small{}$2$}}} & \multicolumn{2}{c}{$-5\,385.710$} &  -5397.683 &  -5356&570 & -5408&260\tabularnewline[-0.65mm]
\multicolumn{2}{c}{\textbf{\textcolor{black}{\small{}$3$}}} & \multicolumn{2}{c}{$-5\,241.269$} &  -5260.427 &  -5280&073 & -5277&492\tabularnewline[-0.65mm]
\multicolumn{2}{c}{\textbf{\textcolor{black}{\small{}$4$}}} & \multicolumn{2}{c}{$-5\,083.976$} &  -5110.717 &  -5138&177 & -5134&567\tabularnewline[-0.65mm]
\multicolumn{2}{c}{\textbf{\textcolor{black}{\small{}$5$}}} & \multicolumn{2}{c}{$-4\,914.852$} & -4980.180 &  -4984&997 & -4980&318\tabularnewline
\hline 
\end{tabular}

\end{table}

\begin{table}
\protect\caption{Quality of MLR$_{6,6}^{6.2}(4)$ fit to original \emph{ab initio }energies
of \cite{Musia2014} for the $3B(3^{1}\Pi_{1_{u}})$. When an isotopologue
listed in Table \ref{tab:3B v-levels} has at least one vibrational
energy level beyond the range of \emph{ab initio }data available,
approximate distances are given for the outer classical turning points
of the corresponding vibrational wavefunctions.\label{tab:3BMLRvsAbInitio}}

\begin{tabular}{cr@{\extracolsep{0pt}.}lcr@{\extracolsep{0pt}.}lr@{\extracolsep{0pt}.}l}
\hline 
\noalign{\vskip2mm}
\textcolor{blue}{\small{}~$r$}\textcolor{blue}{~} & \multicolumn{2}{c}{\textcolor{blue}{\small{}Original }} & \textcolor{blue}{\small{}Weight } & \multicolumn{2}{c}{\textcolor{blue}{\small{}MLR - Original }} & \multicolumn{2}{c}{\textcolor{blue}{\small{}MLR}}\tabularnewline
\textcolor{blue}{{[}$\mbox{\AA}${]}} & \multicolumn{2}{c}{\textcolor{blue}{\small{}{[}cm$^{-1}${]}}} & \textcolor{blue}{\small{}{[}cm$^{-1}${]}} & \multicolumn{2}{c}{\textcolor{blue}{\small{}{[}cm$^{-1}${]}}} & \multicolumn{2}{c}{\textcolor{blue}{\small{}{[}cm$^{-1}${]}}}\tabularnewline[2mm]
\hline 
\noalign{\vskip2mm}
2.223 & -305&82 & 100 & 798&584 & 492&764\tabularnewline[-0.65mm]
2.328 & -1574&16 & 100 & 354&615 & -1219&545\tabularnewline[-0.65mm]
2.434 & -2650&69 & 100 & 133&127 & -2517&563\tabularnewline[-0.65mm]
2.540 & -3531&22 & 100 & 45&218 & -3486&002\tabularnewline[-0.65mm]
2.646 & -4198&64 & 1.00 & 5&965 & -4192&675\tabularnewline[-0.65mm]
2.752 & -4684&78 & 1.00 & -7&168 & -4691&948\tabularnewline
\hline 
\end{tabular}
\end{table}

\begin{table}
\protect\caption{Vibrational energies in cm$^{-1}$ for the $3B(3^{1}\Pi_{1_{u}})$
state predicted by the MLR$_{6,6}^{6.2}(4)$ from Table \ref{tab:MLRpotentials}.
\label{tab:3B v-levels}}

\begin{tabular}{r@{\extracolsep{0pt}.}lr@{\extracolsep{0pt}.}lr@{\extracolsep{0pt}.}lr@{\extracolsep{0pt}.}lr@{\extracolsep{0pt}.}l}
\hline 
\noalign{\vskip2mm}
\multicolumn{2}{c}{\textcolor{black}{\small{}~$v$}} & \multicolumn{2}{c}{\textcolor{blue}{\small{}$^{6,6}$Li$_{2}$}} & \multicolumn{2}{c}{\textcolor{blue}{\small{}$^{6,7}$Li$_{2}$}} & \multicolumn{2}{c}{\textcolor{blue}{\small{}$^{7,7}$Li$_{2}$}} & \multicolumn{2}{c}{\textcolor{blue}{\small{}$^{11,11}$Li$_{2}$}}\tabularnewline[2mm]
\hline 
\noalign{\vskip2mm}
\multicolumn{2}{c}{\textbf{\small{}$\mathfrak{D}_{e}$}} & 5368&8 & 5368&8 & 5368&8 & 5368&8\tabularnewline[-0.65mm]
\multicolumn{2}{c}{\textbf{\textcolor{black}{\small{}$0$}}} & -5259&702 & -5263&694 & -5267&840 & -5267&292\tabularnewline[-0.65mm]
\multicolumn{2}{c}{\textbf{\textcolor{black}{\small{}$1$}}} & -5047&141 & -5058&712 & -5070&746 & -5069&155\tabularnewline[-0.65mm]
\multicolumn{2}{c}{\textbf{\textcolor{black}{\small{}$2$}}} & -4838&570 & -4857&433 & -4877&067 & -4874&470\tabularnewline[-0.65mm]
\multicolumn{2}{c}{\textbf{\textcolor{black}{\small{}$3$}}} & -4634&049 & -4659&912 & -4686&855 & -4683&290\tabularnewline[-0.65mm]
\multicolumn{2}{c}{\textbf{\textcolor{black}{\small{}$4$}}} & -4433&598 & -4466&170 & -4500&132 & -4495&636\tabularnewline[-0.65mm]
\multicolumn{2}{c}{\textbf{\textcolor{black}{\small{}$5$}}} & -4237&212 & -4276&205 & -4316&896 & -4311&508\tabularnewline
\hline 
\end{tabular}

\end{table}

\begin{table}
\protect\caption{Quality of MLR$_{6,8}^{7.0}(8)$ fit to original \emph{ab initio }energies
of \cite{Musia2014} for the $3C(3^{1}\Pi_{1_{g}})$ state. \label{tab:3CMLRvsAbInitio}}

\begin{tabular}{r@{\extracolsep{0pt}.}lr@{\extracolsep{0pt}.}lcr@{\extracolsep{0pt}.}lr@{\extracolsep{0pt}.}l}
\hline 
\noalign{\vskip2mm}
\multicolumn{2}{c}{\textcolor{blue}{\small{}~$r$}\textcolor{blue}{~}} & \multicolumn{2}{c}{\textcolor{blue}{\small{}Original }} & \textcolor{blue}{\small{}Weight } & \multicolumn{2}{c}{\textcolor{blue}{\small{}MLR - Original }} & \multicolumn{2}{c}{\textcolor{blue}{\small{}MLR}}\tabularnewline
\multicolumn{2}{c}{\textcolor{blue}{{[}$\mbox{\AA}${]}}} & \multicolumn{2}{c}{\textcolor{blue}{\small{}{[}cm$^{-1}${]}}} & \textcolor{blue}{\small{}{[}cm$^{-1}${]}} & \multicolumn{2}{c}{\textcolor{blue}{\small{}{[}cm$^{-1}${]}}} & \multicolumn{2}{c}{\textcolor{blue}{\small{}{[}cm$^{-1}${]}}}\tabularnewline[2mm]
\hline 
\noalign{\vskip2mm}
 2&328 &   -331&50 &  10.00 &  35&693 & -295&807\tabularnewline[-0.65mm]
 2&434 &  -1429&53 &  10.00 &  -7&764 & -1437&294\tabularnewline[-0.65mm]
  2&540 &  -2276&70 &  10.00 & -22&184 & -2298&884\tabularnewline[-0.65mm]
   2&645 &  -2916&47 &  10.00 & -20&557 & -2937&027\tabularnewline[-0.65mm]
   2&751 &  -3385&05 &   5.00 & -11&667 & -3396&717\tabularnewline[-0.65mm]
   2&857 &  -3712&29 &   5.00 &  -1&457 & -3713&747\tabularnewline
\hline 
\end{tabular}
\end{table}

\begin{table}
\protect\caption{Vibrational energies in cm$^{-1}$ for the $3C(3^{1}\Pi_{1_{g}})$
state predicted by the MLR$_{6,8}^{7.0}(8)$ from Table \ref{tab:MLRpotentials}.\label{tab:3C v-levels}}

\begin{tabular}{r@{\extracolsep{0pt}.}lr@{\extracolsep{0pt}.}lr@{\extracolsep{0pt}.}lr@{\extracolsep{0pt}.}lr@{\extracolsep{0pt}.}l}
\hline 
\noalign{\vskip2mm}
\multicolumn{2}{c}{\textcolor{black}{\small{}~$v$}} & \multicolumn{2}{c}{\textcolor{blue}{\small{}$^{6,6}$Li$_{2}$}} & \multicolumn{2}{c}{\textcolor{blue}{\small{}$^{6,7}$Li$_{2}$}} & \multicolumn{2}{c}{\textcolor{blue}{\small{}$^{7,7}$Li$_{2}$}} & \multicolumn{2}{c}{\textcolor{blue}{\small{}$^{11,11}$Li$_{2}$}}\tabularnewline[2mm]
\hline 
\noalign{\vskip2mm}
\multicolumn{2}{c}{\textbf{\small{}$\mathfrak{D}_{e}$}} & \multicolumn{2}{c}{} & \multicolumn{2}{c}{} & \multicolumn{2}{c}{} & \multicolumn{2}{c}{}\tabularnewline[-0.65mm]
\multicolumn{2}{c}{\textbf{\textcolor{black}{\small{}$0$}}} &  -3957&890 & -3961&862 &  -3965&987 & -3965&442\tabularnewline[-0.65mm]
\multicolumn{2}{c}{\textbf{\textcolor{black}{\small{}$1$}}} &  -3745&119 & -3761&862 &  -3765&987 & -3765&442\tabularnewline[-0.65mm]
\multicolumn{2}{c}{\textbf{\textcolor{black}{\small{}$2$}}} &  -3445&119 & -3552&904 &  -3572&894 & -3570&251\tabularnewline[-0.65mm]
\multicolumn{2}{c}{\textbf{\textcolor{black}{\small{}$3$}}} &  -3324&020 & -3350&648 &  -3378&354 & -3374&690\tabularnewline[-0.65mm]
\multicolumn{2}{c}{\textbf{\textcolor{black}{\small{}$4$}}} &  -3116&570 & -3150&393 &  -3185&613 & -3180&954\tabularnewline[-0.65mm]
\multicolumn{2}{c}{\textbf{\textcolor{black}{\small{}$5$}}} &  -2911&754 & -2952&521 &  -2995&010 & -2989&387\tabularnewline
\hline 
\end{tabular}

\end{table}

\begin{table}
\protect\caption{Quality of MLR$_{6,9}^{7.2}(4)$ fit to original \emph{ab initio }energies
of \cite{Musia2014} for the $3d(3^{3}\Pi_{2_{g}})$ state. When an
isotopologue listed in Table \ref{tab:3d v-levels} has at least one
vibrational energy level beyond the range of \emph{ab initio }data
available, approximate distances are given for the outer classical
turning points of the corresponding vibrational wavefunctions.\label{tab:3dMLRvsAbInitio}}

\begin{tabular}{r@{\extracolsep{0pt}.}lr@{\extracolsep{0pt}.}lcr@{\extracolsep{0pt}.}lr@{\extracolsep{0pt}.}l}
\hline 
\noalign{\vskip2mm}
\multicolumn{2}{c}{\textcolor{blue}{\small{}~$r$}\textcolor{blue}{~}} & \multicolumn{2}{c}{\textcolor{blue}{\small{}Original }} & \textcolor{blue}{\small{}Weight } & \multicolumn{2}{c}{\textcolor{blue}{\small{}MLR - Original }} & \multicolumn{2}{c}{\textcolor{blue}{\small{}MLR}}\tabularnewline
\multicolumn{2}{c}{\textcolor{blue}{{[}$\mbox{\AA}${]}}} & \multicolumn{2}{c}{\textcolor{blue}{\small{}{[}cm$^{-1}${]}}} & \textcolor{blue}{\small{}{[}cm$^{-1}${]}} & \multicolumn{2}{c}{\textcolor{blue}{\small{}{[}cm$^{-1}${]}}} & \multicolumn{2}{c}{\textcolor{blue}{\small{}{[}cm$^{-1}${]}}}\tabularnewline[2mm]
\hline 
\noalign{\vskip2mm}
 2&222544 &   -675&63 & 150.00 &  242&061 & -433&569\tabularnewline[-0.65mm]
 2&328380 &  -2176&84 & 15.00 &   96&978 & -2079&862\tabularnewline[-0.65mm]
 2&434215 &  -3353&22 & 10.00 &   27&796 & -3325&424\tabularnewline[-0.65mm]
 2&540051 &  -4256&36 &  1.00 &    1&853 & -4254&507\tabularnewline[-0.65mm]
 2&645886 &  -4929&71 &  1.00 &   -3&720 & -4933&430\tabularnewline[-0.65mm]
 2&751721 &  -5412&99 &  1.00 &   -1&158 & -5414&148\tabularnewline
\hline 
\end{tabular}
\end{table}

\begin{table}
\protect\caption{Vibrational energies in cm$^{-1}$ for the $3d(3^{3}\Pi_{2_{g}})$
state predicted by the MLR$_{6,9}^{7.2}(4)$ from Table \ref{tab:MLRpotentials}.\label{tab:3d v-levels}}

\begin{tabular}{r@{\extracolsep{0pt}.}lr@{\extracolsep{0pt}.}lr@{\extracolsep{0pt}.}lr@{\extracolsep{0pt}.}lr@{\extracolsep{0pt}.}l}
\hline 
\noalign{\vskip2mm}
\multicolumn{2}{c}{\textcolor{black}{\small{}~$v$}} & \multicolumn{2}{c}{\textcolor{blue}{\small{}$^{6,6}$Li$_{2}$}} & \multicolumn{2}{c}{\textcolor{blue}{\small{}$^{6,7}$Li$_{2}$}} & \multicolumn{2}{c}{\textcolor{blue}{\small{}$^{7,7}$Li$_{2}$}} & \multicolumn{2}{c}{\textcolor{blue}{\small{}$^{11,11}$Li$_{2}$}}\tabularnewline[2mm]
\hline 
\noalign{\vskip2mm}
\multicolumn{2}{c}{\textbf{\small{}$\mathfrak{D}_{e}$}} & \multicolumn{2}{c}{\textcolor{black}{$\;\;6\,045.135$~~}} & \multicolumn{2}{c}{\textcolor{black}{$\;\;6\,045.135$~~}} & \multicolumn{2}{c}{\textcolor{black}{$\;\;6\,045.135$~~}} & \multicolumn{2}{c}{\textcolor{black}{$\;\;6\,045.135$~~}}\tabularnewline[-0.65mm]
\multicolumn{2}{c}{\textbf{\textcolor{black}{\small{}$0$}}} &  -5909&082 &  -5914&058 & -5919&226 &  -5918&543\tabularnewline[-0.65mm]
\multicolumn{2}{c}{\textbf{\textcolor{black}{\small{}$1$}}} &  -5641&796 &  -5656&375 & -5671&532 &  -5669&528\tabularnewline[-0.65mm]
\multicolumn{2}{c}{\textbf{\textcolor{black}{\small{}$2$}}} &  -5377&524 &  -5401&482 & -5371&532 &  -5369&528\tabularnewline[-0.65mm]
\multicolumn{2}{c}{\textbf{\textcolor{black}{\small{}$3$}}} &  -5116&452 &  -5149&547 & -5183&994 &  -5179&438\tabularnewline[-0.65mm]
\multicolumn{2}{c}{\textbf{\textcolor{black}{\small{}$4$}}} &  -4858&751 &  -4949&547 & -4944&436 &  -4938&652\tabularnewline[-0.65mm]
\multicolumn{2}{c}{\textbf{\textcolor{black}{\small{}$5$}}} &  -4604&574 &  -4655&149 & -4707&858 &  -4700&882\tabularnewline
\hline 
\end{tabular}

\end{table}

\clearpage{}

\end{document}